\documentclass[1pt,a4paper]{article}

\usepackage[a4paper]{geometry}
\usepackage[utf8]{inputenc}
\usepackage[english]{babel}

\usepackage{cite}
\usepackage{url}
\usepackage[breaklinks]{hyperref}

\usepackage{graphicx}
\usepackage{caption}
\usepackage{csquotes}

\usepackage{diagbox}
\usepackage{amssymb}
\usepackage{multirow}
\usepackage{enumitem}
\usepackage{cleveref}

\newcommand{\keywords}[1]{{\small \textbf{\textit{Keywords}}--- #1}}

\begin{document}
	\title{Patterns for Blockchain Data Migration}
	\author{HMN Dilum Bandara and Xiwei Xu \\
		Data61, CSIRO, Sydney, Australia \\
		firstname.lastname@data61.csiro.au \\ \\
		Ingo Weber \\
		Chair of Software and Business Engineering \\
		Technische Universitaet Berlin, Berlin, Germany \\
		firstname.lastname@tu-berlin.de
	}
	\date{}
	\maketitle
	
	\begin{abstract}
With the rapid evolution of technological, economic, and regulatory landscapes, contemporary blockchain platforms are all but certain to undergo major changes. Therefore, the applications that rely on them will eventually need to migrate from one blockchain instance to another to remain competitive and secure, as well as to enhance the business process, performance, cost efficiency, privacy, and regulatory compliance. However, the differences in data and smart contract representations, modes of hosting, transaction fees, as well as the need to preserve consistency, immutability, and data provenance introduce unique challenges over database migration. We first present a set of blockchain migration scenarios and data fidelity levels using an illustrative example. We then present a set of migration patterns to address those scenarios and the above data management challenges. Finally, we demonstrate how the effort, cost, and risk of migration could be minimized by choosing a suitable set of data migration patterns, data fidelity level, and proactive system design. Practical considerations and research challenges are also highlighted.
	\end{abstract}

 	
 	 	\vfill\fbox{\parbox{\dimexpr\columnwidth-15pt}{
 			Published paper available at: \url{https://doi.org/10.1145/3424771.3424796}

 			Citation: H.M.N. Dilum Bandara, Xiwei Xu, and Ingo Weber, ``Patterns for Blockchain Data Migration,'' European Conf. on Pattern Languages of Programs 2020 (EuroPLoP 2020), July 2020, Virtual Event, Germany. doi: 10.1145/3424771.3424796.
 	}} 
 
	\section{Introduction}\label{intro}
Since the launch of Bitcoin over a decade ago~\cite{Nakamoto08}, an unprecedented number of blockchain platforms with different designs, features, and operational models have emerged. While each claims its superiority over predecessors in terms of performance, features, security, or governance, given an application scenario, it is non-trivial to identify a fitting blockchain platform. However, not wanting to lose the early adopter advantage, even 
enterprise information systems and business process management systems are starting to adopt blockchain platforms~\cite{Bizagi8, Pintado19}. Alternatively, as the technological, business, economic, and regulatory landscapes are still evolving, it is quite unclear what blockchain platforms will make the cut in 
such domains. 
Therefore, if the chosen blockchain turned out to be ill-suited, it might be difficult, costly, and risky to change it. This is due to the incompatibilities in platforms, mode of hosting, and blockchain properties such as consistency, immutability, transparency, and openness. Thus, even though data migration has been an afterthought, it is imperative to know the feasibility and caveats of blockchain migration.

\begin{figure}[htb!]
	\centering
	\includegraphics[width=0.6\textwidth]{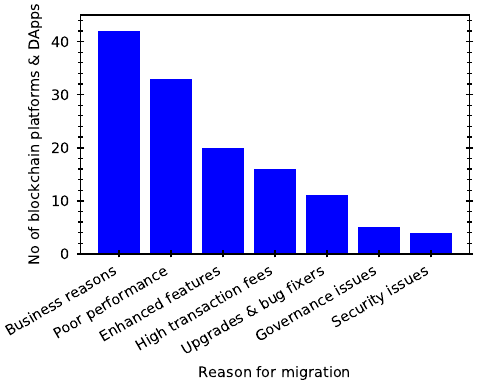}
	\caption{Reasons for blockchain data migration.\label{fig:reason}}
\end{figure}

An application that uses a blockchain as an underlying data store may opt to migrate its data due to diverse reasons. For example, Fig. \ref{fig:reason} shows the reason for data migration as cited by 72 blockchain platforms and Decentralized Applications (DApps) that migrated their data between July 2017 and April 2020 (see \Cref{patterns_overview} for details). Business reasons include the interest to launch own blockchain instances, partnerships, mergers and acquisitions, and multi-blockchain operations. Another key reason is the emergence of new blockchain platforms with better performance (i.e., higher throughput, lower latency, or faster finality), new features, and low transaction fees compared to incumbent platforms such as Bitcoin and Ethereum. Essential upgrades due to the blockchain platform changes, bug fixes, security, and governance issues also lead to migration. For example, Gartner expects that ``through 2021, 90\% of the enterprise blockchain implementations will require replacement within 18 months to remain competitive and secure, and to avoid obsolescence''~\cite{Heudecker18}.  Most reasons for migration stem from the immaturity of technical, business, economic, and regulatory facets of blockchains. However, in the same way that database migration and enterprise application integration have not gone away, data migration in the blockchain context is also a lasting problem. For example, business mergers and acquisitions, establishing/joining a new consortium in cross-organizational processes, as well as regulatory changes~\cite{Hive20, Morris12} may also force an organization to move into a consortium blockchain or \textit{Blockchain as a Service} (BaaS) platform. Moreover, business process reengineering~\cite{Mendling2018}, separation of internal and shared data, change of hardware, and consolidation, may also trigger a migration~\cite{Morris12}. Furthermore, an organization may also adopt multiple blockchains with different cost, performance, and workload characteristics, which might involve partial data migration.

While most applications use blockchain as a data store, blockchains have several notable differences compared to conventional databases. For example, a blockchain could be abstracted as a \textit{key-value} store that maintains a set of states ~\cite{Paik2019}. Thus, the schema-less nature makes the mapping between blockchains relatively easy. While blockchain transactions could perform complex operations on multiple states, they do not fully support CRUD operations and ACID properties \cite{Paik2019, Tai17}. Nevertheless, the quality of data on a blockchain is generally high due to the data consistency and completeness provided by consensus and immutability, respectively. Optionally, blockchains might include business logic in the form of smart contracts ~\cite{Wood19}. Smart contracts are not only more expressive and complex than stored procedures but also specific to a blockchain platform and instance~\cite{Amani18}. Thus, smart contracts may need to be ported to the target blockchain leading to potential errors despite significant time and cost needed to modify/rewrite and test them. Moreover, as smart contracts have embedded data, ported smart contracts also need to reestablish their data. However, the global state, transaction sequence, and history cannot be arbitrarily recreated on the target blockchain due to the consensus process and the need to preserve consistency, transparency, and data provenance. Even when they are recreated on the target blockchain, records must be kept on the blockchain itself on how the new data came into existence and any changes made to data during the migration process. Furthermore, as transactions are digitally signed, they may not be replayable across blockchains without having access to the private keys. Also, blockchain migration has to be a single-shot process, as rollbacks could be impossible due to immutability. The replication factor of blockchains is several orders of magnitude higher than in databases, and both the replica holders and users have a say in the governance process. Thus, in addition to paying transaction fees to offset the cost of resource utilization, multi-party approval is needed to introduce any changes to data and platform. Therefore, even when workarounds are possible, significant time and cost are required to migrate a large volume of accounts, native assets, states, transactions, and smart contracts that are interrelated. Consequently, blockchain migration becomes a nontrivial, costly, and risky process~\cite{Heudecker18} compared to database migration.  While we can learn from database migration best practices~\cite{Morris12, Kim18} and numerous patterns~\cite{Wagner11, Rocha15, Ruping13, Strauch13}, it is imperative to answer the following research question to understand the extent to which we can generalize the related work and identify new patterns to address unique challenges of blockchains:

\medskip
{\centering \textit{What are the effective patterns for safeguarding data 
		when migrating from one blockchain to another}? \par}
\medskip

In this paper, we first explain a set of blockchain migration scenarios and data fidelity levels using an illustrative example. The example is derived based on the literature and our experience from numerous blockchain projects for industry and government agencies. Second, we identified eleven patterns to achieve those migration scenarios under varying data fidelity levels. While four of these patterns stem from database migration, specific adaptations are needed to support blockchain migration. Three other patterns from blockchain and database literature are also adopted to address non-functional aspects such as quality of migrated data, cost, and privacy. Third, using the illustrative example, we discuss how the proposed patterns and data fidelity levels could be used to address the identified migration scenarios and data management challenges. The proposed patterns are applicable to software architects, developers, system administrators, and project leads (both technical and business) who need to plan, develop, configure, and monitor blockchain data migration projects. In conclusion, while migrating to a private or consortium blockchain could be achieved relatively easily, recreating full blockchain history on an existing public blockchain is impractical. Nevertheless, the global state can still be recreated, which is sufficient for most practical migration scenarios. Therefore, the success of blockchain migration boils down to choosing a suitable data fidelity level and a set of data migration patterns that balances competing factors such as performance, cost, time, effort, granularity of data, transparency, security, privacy, and risk. Finally, we discuss practical considerations and several research challenges.

The rest of the paper is organized as follows. \Cref{back} defines terminology and presents an illustrative example. Migration scenarios and data fidelity levels are presented in \Cref{scenarios}, and patterns are proposed in \Cref{patterns}. Migration pattern to scenario mapping and a set of use cases are presented in \Cref{mapping}. \Cref{discuss} discusses practical considerations and research challenges. Concluding remarks are highlighted in \Cref{summary}.

\section{Preliminaries}\label{back}

\subsection{Background and Definitions}\label{def}
We first define a set of terms related to blockchain data and migration. While most of these terms originate from Ethereum and Bitcoin, they are widely adopted by the blockchain industry and research community \cite{Hyperledger19, EOS18, Hearn16}. These terms are defined to clearly express the illustrative example, migration scenarios, data fidelity levels, and proposed patterns.

A \textit{state} is anything tracked by a blockchain such as the balance of an account, Unspent Transaction Output (UTXO), value of an asset, ownership of an asset represented as a digital token, or an attribute of a physical object. Public blockchains use one or more native assets (i.e., default cryptocurrency) for value transfer and transaction fees. 
Our definition of state also includes the states embedded within a smart contract such as tokens and other data. An \textit{account} (aka., address) is a reference/key to a state or smart contract, e.g., owner of a UTXO in Bitcoin \cite{Nakamoto08} or balance and other data of an Ethereum account \cite{Wood19}. \textit{Global state} (aka., world state) is the collection of all accounts and their current states tracked by the blockchain. 

A \textit{transaction} (TX) is used to initiate a valid state transition. For example, a transaction can change the ownership of a title or debit cryptocurrency from one account and credit it to another. A \textit{Smart Contract} (SC) is a set of executable instructions that are activated in response to a transaction. When executing, these instructions may change the states and call other functions on the blockchain. 
\textit{History} (aka., block history) is the collection of all blocks produced by the blockchain. Each block tracks all the included transactions and resulting states after executing those transactions and smart contracts. 

\textit{Blockchain platform} is the software needed to operate a blockchain \cite{2019Xu}, such as the client software for Bitcoin, Ethereum, or Hyperledger. Execution of such platform software with a specific set of nodes, storage, permissions, and configurations is referred to as a \textit{blockchain instance}. We consider scenarios where an application using a blockchain instance should be either move to a different instance or hardware. The current blockchain instance is referred to as the \textit{source blockchain} and the one after the change is referred to as the \textit{target blockchain}. The source and target blockchains could be on the same or different instances with varying platforms and levels of permissions such as private, consortium, and public.

Different terms such as migration, transfer, conversion, moving, and replication refer to the processes of copying data from a source data store to the target. Moreover, blockchain-specific terminology such as mainnet migration/swap \cite{Arron19, Richard18}, token migration/swap \cite{Vechain18, Milano18}, DApp migration/swap, smart contract migration, bootstrapping \cite{Peter14, Garoffolo18}, hard spooning \cite{Bithereum18}, and teleportation \cite{Sigman} also refer to various forms of data copying within and across blockchains. Therefore, in this paper, we define \textit{blockchain data migration} as the process of recreating full or part of the accounts, states, transactions, smart contracts, and history on the target blockchain while adhering to the consistency, immutability, and transparency properties of blockchains. Underneath, this not only includes data transfer, but also distributed commutation, data exchange, and consensus across a large number of distributed nodes.

Migration issues may also arise within the scope of a single blockchain platform, e.g., across versions, smart contract languages, and APIs. For example, some smart contracts and DApps had to be upgraded when Ethereum Virtual Machine (EVM) instructions and their gas consumption were changed to strengthen the security and optimize the cost and performance \cite{Kyber19}. Similarly, Ethereum smart contracts written in Serpent language need to be rewritten, as Serpent is now deprecated due to known weaknesses \cite{Augur17}. Also, Ethererum Web3 API has many upgraded and depreciated functions, in turn necessitating DApp upgrades to remain relevant. The above definition of blockchain data migration includes scenarios where data need to be recreated on the same or a different blockchain instance. However, we exclude reconfiguration, software upgrade, and cross-chain operation where data remain backward compatible and within the same blockchain instance. Such examples include changes to DApps and smart contracts that do not require movement of data, soft forks where old blocks remain valid under updated state transition rules, and changing the consensus algorithm in Hyperledger Fabric while retaining the state database \cite{Hyperledger19}. While we also consider migration as a one-way process with no return \cite{Morris12}, we consider the two options of continuing or decommissioning the source blockchain after the migration. Such consideration is necessary as a public blockchain may continue, even though an alt-coin (i.e., alternative currency) moves to its blockchain instance \cite{Sedgwick19, Qubicles19, Garoffolo18}.

\subsection{Motivating Application Scenario}\label{example}

Based on the reasons for blockchain data migration listed in Fig. \ref{fig:reason} and our experience in multiple blockchain projects, we construct a hypothetical scenario to illustrate different data migration scenarios, migration options, proposed patterns, and their interrelationships. This scenario captures the growing concerns on poor performance, increasing transaction fees, and poor quality of service experienced in public blockchains, like Ethereum \cite{BitGuild18}, NEO \cite{Effect.AI19}, EOS \cite{KARMA19}, and IOTA \cite{IOTA20}.

Suppose a nonprofit hosted a concert to raise funds for its charity work. The nonprofit pledged to allocate at least 50\% of the profit to the charity work. The nonprofit puts \$5,000 as the seed money, hoping to recover it from the remaining profit. It further planned to set aside up to \$3,000 as next year's seed money and use the remaining balance for further charity work. Moreover, they reached out to several sponsors and donors to raise additional capital. Necessary expenses include payments to key artists (although most artists volunteered), venue, equipment, and marketing. 

The nonprofit decided to use a blockchain, as it hoped better transparency in managing funds would attract more attendees, sponsors, and donors. Pledge to charity work and recovery of seed money were defined as smart contracts. In case of a loss, the nonprofit planned to settle part of the dues to the venue and equipment from a future event. The nonprofit chose a public blockchain that does not charge transaction fees. A budget was prepared, and separate accounts were created for each budget item, and transaction records were maintained using the blockchain. A \textit{ConcertCoin} was defined to minimize the impact of cryptocurrency fluctuation and enhance the traceability. Several exchanges also agreed to buy and sell ConcertCoins without a commission. Seed money was then converted to ConcertCoin and dispersed to accounts using a funding smart contract. Another set of smart contracts was defined to disperse the funds, as well as to enforce budget limits, spending rules, and payments.

The concert was an overwhelming success, partly due to its high transparency in handling finances and ensuring that committed funds do go to the charity work. Eventually, the ledger was balanced, and all the payments, seed money, and next-year concert's retention were settled as per the smart contracts. Given the overwhelming success of the concert and greater transparency it demonstrated, the nonprofit planned on using the blockchain for other fundraisers and to track the disbursement of aid. 

However, soon after, the blockchain community 
started charging transaction fees, as it was grappled with spam transactions. Consequently, exchanges were also forced to charge a commission for ConcertCoin conversions. It was further realized that the cost and performance problems are unlikely to be solved soon, as changes to the blockchain platform's architecture and incentive mechanism are needed to fix the issue. Given these circumstances and the need to integrate future fundraisers and internal activities, the nonprofit decided to change the blockchain instance.

In this case, the nonprofit has several options, such as moving to a private or another public blockchain. Under the first option, the nonprofit could use the same or a different blockchain platform from a BaaS provider or host its instance on-premise or on the cloud. The nonprofit also has a range of choices when it comes to migrating the data to the new blockchain instance, e.g., starting the application without any history of the concluded concert, with the closing balances, or with the entire history.

\section{Blockchain Migration Scenarios}\label{scenarios}

We consider different choices for the target blockchain instance, such as preexisting or new; private, consortium, or public; and same or a different blockchain platform. We also consider the possibility of not decommissioning the source blockchain after the migration. However, we do not attempt to recommend a target blockchain, and instead assume that the nonprofit has already chosen a target blockchain based on competing factors such as performance, cost, time, effort, granularity of data, security, and privacy. The nonprofit has a few options such as to cash out and start anew, transfer only the closing balance(s), or transfer the global state and history to the target blockchain. In the context of the nonprofit’s migration requirements, following data migration scenarios can be identified:

\begin{enumerate}
	\item 
	\textit{Relocate} – changes the hardware on which the blockchain instance runs. For example, the nonprofit may run the same blockchain platform as an on-premise or cloud (IaaS or BaaS) instance to gain cost efficiency, performance, and privacy.
	
	\item \textit{Upgrade} – creates a new blockchain or smart contract instance to gain better performance, security, privacy, novel features, and lower cost while losing backward compatibility. For example, the nonprofit may move from proof of work to proof of stake version of the same blockchain platform to get better performance and cost-efficiency \cite{Gifto19, Milano18}.
	
	\item \textit{Consolidate} – combines two or more existing blockchains to a single target blockchain. For example, the concert application could move to another existing blockchain to get benefits similar to upgrading.
	
	\item 
	\textit{Separation} – forks off one or more target blockchains and partitions the global state across them. For example, the concert application running on the public blockchain can declare independence by forking off its instance and managing all its states on the new instance.
	
	\item \textit{Archive} – creates a full archive of block history on the target blockchain. For example, the nonprofit may create an archive, as the source blockchain prunes old blocks to reduce storage or go out of business. Moreover, the archive can be used to serve read-only workloads targeting transaction validation, auditing, data analytics, and public access.
\end{enumerate}	

Differences between the source and target blockchain platforms, their existence, and mode of deployment determine what and how data can be extracted from the source and recreated on the target. Moreover, as per the third golden rule of Morris's data migration \cite{Morris12}, ``no one needs, wants, or will pay for perfect data.'' However, in the context of blockchains, the absolute correctness of asset ownership, data provenance, and smart contracts are essential to gain complete trust. Also, transaction history may need to be retained for years, e.g., the banking and finance industry typically retains transaction history for five to seven years. Therefore, given a migration scenario, it is imperative to identify what data to migrate and what to retain on the source blockchain while finding the right balance among blockchain constraints; differences between platforms, their existence, and mode of deployment; preserving trust, auditability, and compliance; data consistency, granularity, and utility; and cost and time to migrate. Thus, while moving to the target blockchain, the nonprofit may decide to migrate data at different levels of fidelity as follows:

\begin{enumerate}
	\item \textit{Fresh Start} – Start a new application round/instance without migrating any blockchain data. For example, the nonprofit could cash out any remaining ConcertCoins and disregard blockchain state, smart contracts, and history on the source blockchain. Next year, it can start the concert application anew on a different blockchain instance.
	
	\item \textit{State Only} – Migrate a chosen subset of states that is essential to transact during the next application round to the target blockchain. For example, the nonprofit could transfer next year’s seed money to the new blockchain instance. It may also transfer the closing balance of each budget item when commitments to vendors/artists are pending, or spending rules are attached to budget items.
	
	\item \textit{State and Transactions} – Migrate both the chosen states and associated transactions. For example, in addition to recreating closing balance(s), it may be necessary to look-up past transactions on the target blockchain.
	
	\item \textit{Genesis and Transactions} – Initial states and all subsequent transactions are migrated to the target blockchain to recreate the global state and full history. This enables the nonprofit to preserve the integrity of the data and facilitate auditing.
	
	\item \textit{Blockchain History} – Full history of the source blockchain (including state, transactions, smart contracts, and blocks) are migrated to the target blockchain. This enables the nonprofit to facilitate transparency, auditing, and data analytics.
	
\end{enumerate}

As we go down the levels, completeness of the data migrated to the target blockchain increases. All related smart contracts and their embedded data need to be migrated as per the selected fidelity level.

\section{Blockchain Migration Patterns}\label{patterns}
\subsection{Migration Architecture and Pattern Overview}\label{patterns_overview}

Many different constraints need to be overcome while migrating data under the above scenarios and fidelity levels. While each migration scenario is unique, we can always learn from the commonly occurring problems and recurring solutions to those. These reusable solutions can be formulated as a set of technology and implementation-independent \textit{patterns}. Patterns demonstrate techniques and strategies to meet the requirements of a blockchain migration project while minimizing risk and cost.

Data migration patterns are derived from our experience in multiple blockchain projects and related work. While database migration has been discussed in the formal literature, details on blockchain migration are available only in grey literature (e.g., white papers, online news sites, blogs, and videos). Therefore, we used a Multivocal Literature Review (MLR) \cite{Garousi2019} -- a form of systematic literature review that includes the grey literature -- as our methodology to explore the reasons and techniques for blockchain data migration. MLR process started with our research question in \Cref{intro}. We used Google Scholar, News, and Search as the data sources. Search keywords included ``blockchain,'' ``DApp,'' ``token,'' ``smart contract,'' ``migration,'' ``swap,'' ``mainnet,'' and their combinations. Snowball sampling was used to expand the pool of potential sources by following citations and links from already found sources. The process was repeated until theoretical saturation was reached. Then the pool of sources was filtered according to predefined inclusion and exclusion criteria, e.g., ``migration must include movement of data,'' ``must specify reasons and techniques for migration,'' ``and migration must be completed''. Information sources from the blockchain platform and DApp governance body were given higher preference over third-party sources. Key formal literature in the vetted pool of sources included smart contract migration \cite{Gueta19, Amani18}, blockchain patterns \cite{Xu18, 2019Xu}, database migration and patterns \cite{Morris12, Ruping13, Rocha15, Wagner11, Kim18, Haller09}, and cloud patterns \cite{Strauch13}. While 128 web pages related to blockchain and DApp migration were selected, only the ones related to the presented patterns and with a detailed explanation of the migration process are listed in the paper \cite{Aeternity19, Arron19, Augur17, Bitcoin18, BitcoinWiki, BitGuild18, Bithereum18, Buterin17, Counterparty, Effect.AI19, EOS18, Garoffolo18, Gifto19, Hearn16, Hive20, Hyperledger18, IOTA20, KARMA19, Kin18, Kyber19, Leonowicz19, Mahalingam18, Milano18, Moonlight19, Peter14, Qubicles19, Richard18, Sedgwick19, SFOX19, Sigman, Storj17, Telos18, Thomas15, Tomo18, TRON18, Tronbet19, Vechain18, Wilcke16, Wright20, Zafar18, Zeepin18, Zhang19}. Finally, the vetted pool of sources was analyzed using open and axial coding. This process was iterated several times to derive a set of generic and related patterns. Consequently, we identified reasons (see Fig. \ref{fig:reason}) and techniques used by 72 cases of blockchain and DApp data migrations between July 2017 and April 2020.

\begin{figure}[htb!]
	\centering
	\includegraphics[width=0.85\textwidth]{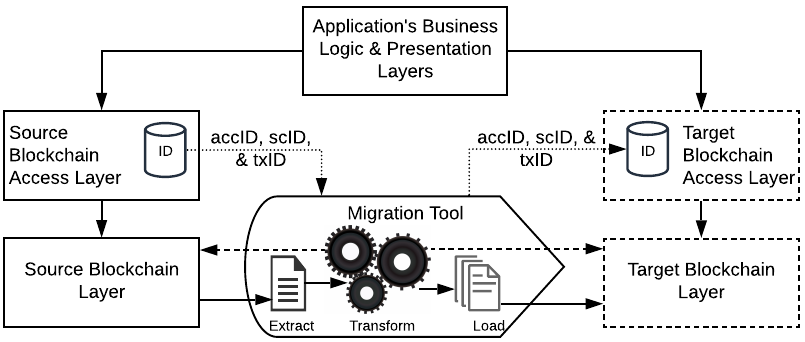}
	\caption{Data migration architecture.\label{fig:archi}}
\end{figure}

We explain the patterns in the context of data migration architecture illustrated in Fig. \ref{fig:archi}. Similar to database migration, we envision the migration team will utilize a tool (either developed in-house or off-the-shelf) to simplify the migration process. The migration tool could follow the Extract, Transform, and Load (ETL) process \cite{Ruping13, Haller09} to copy data from the source blockchain and recreate them on the target blockchain. However, phases of the ETL process may be interchanged as per the chosen set of migration patterns. For example, a state could be transformed as a token on the source blockchain before extraction. Moreover, to preserve the consistency and accountability, any data transformation must be recorded on either the source or target blockchain. Thus, most transformations happen within a blockchain rather than in a separate staging area. Scattered arrows in the figure indicate this. 

Due to the incompatibilities between the source and target blockchains' data representations, as well as the creation of new accounts, smart contracts, and replay of transactions, changes may be needed at the Blockchain Access/API Layer (BAL) \cite{Paik2019, Strauch13}. Similar to the data access layer in databases, BAL abstracts the connectivity to the blockchain. It may also map application-level references to blockchain identifiers (ID) as they are very different. For example, a username used by a DApp needs to be mapped to the user's address or public key on the blockchain. Such application-level reference to blockchain ID mapping is usually maintained in a protected database within BAL, which we refer to as the \textit{ID database}. When the application holds the user's private key (e.g., custodial wallet), keys may also be maintained in this database. Therefore, in addition to updating the BAL to integrate the target blockchain, ID database within the BAL needs to be updated to reflect new account and smart contract addresses, keys, and transaction IDs during the migration. Moreover, ID database can be used to identify what accounts, states, transactions, and smart contracts to migrate, as blockchains try to be anonymous by not keeping track of applications and their users. Furthermore, even the migration tool could use the BAL to access both the blockchains. Therefore, BAL and its ID database are an integral part of the migration architecture. The dotted lines in Fig. \ref{fig:archi} show the flow of account (\texttt{accID}), smart contract (\texttt{scID}), and transaction (\texttt{txID}) identifiers from/to ID database. 

\begin{figure*}[htb!]
	\centering
	\includegraphics[width=\textwidth]{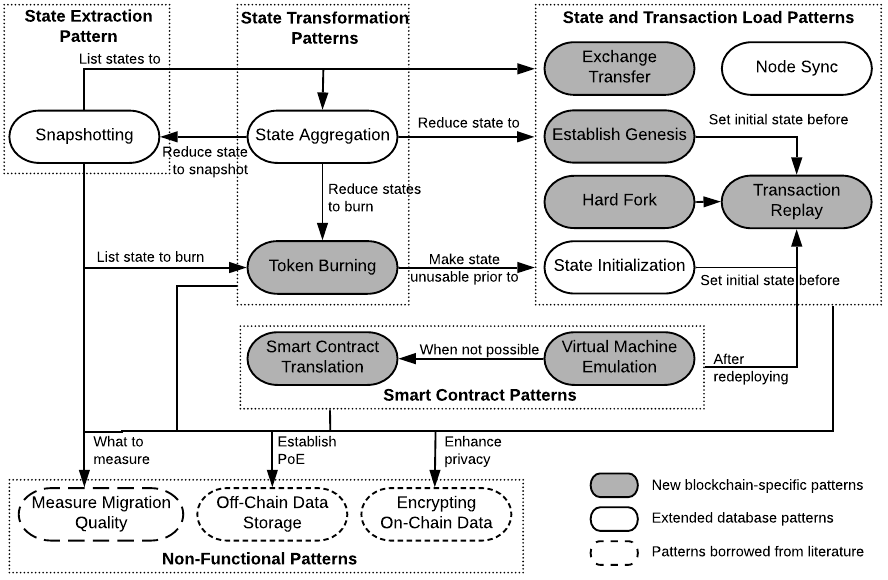}
	\caption{Overview of patterns.\label{fig:patterns}}
\end{figure*}

Fig. \ref{fig:patterns} shows the proposed blockchain migration patterns, grouping, and relationships. We grouped the patterns used to copy data between source and target blockchains based on the phases in the ETL process. Patterns related to smart contracts and non-functional aspects are also arranged into respective groups. For example, the \textit{snapshotting} pattern in the \textit{state extraction} pattern group is used to extract states from the source blockchain. States can be transformed before or after the extraction using \textit{state transformation} patterns. For example, the \textit{state aggregation} pattern can combine multiple states to reduce the volume of data to be migrated. States could be marked as unusable on the source blockchain using the \textit{token burning} pattern. Moreover, \textit{snapshotting} could also be used to specify the list of states to be aggregated or marked as unusable. Patterns in the \textit{state and transaction load} group can be applied to recreate load states and transactions on the target blockchain. For example, the \textit{establish genesis} pattern can be used to spin up a new blockchain instance by including the extracted data in the genesis block. In contrast, the extracted data can be appended to an existing blockchain using the \textit{hard fork} pattern. \textit{Exchange transfer} pattern can be used to transfer tradable assets/states between blockchains, whereas \textit{state initialization} pattern allows arbitrary reaction of other forms of states. \textit{Transaction replay} pattern can be used to replay transactions on the target blockchain. \textit{Node sync} pattern can establish the full history on the target blockchain. \textit{Smart contract} patterns present techniques to migrate smart contracts either by reusing the smart contract execution environment on the target blockchain or translating smart contracts. \textit{Smart contract} patterns need to be used before recreating the states using \textit{state and transaction load} patterns. 

\textit{Non-functional} patterns could assist other pattern groups to optimize storage, enhance security, and measure the success of migration. For example, it is expensive and impractical to store a large volume of data on the target blockchain. However, when the source blockchain is decommissioned, we still need to store large snapshots, details of state aggregation, as well as updates to the ID database and smart contracts on the target blockchain to preserve accountability, transparency, and integrity. In such cases, the \textit{off-chain data storage} pattern \cite{Xu18} could be used to add a Proof of Existence (PoE) entry by adding a hash of the required data to the blockchain while storing the actual data offline. While migrating sensitive data to a public blockchain, \textit{encrypting on-chain data} pattern \cite{Xu18} could be used to encrypt the states and transaction data to enhance the data confidentiality. The \textit{measure migration quality} pattern \cite{Wagner11} uses metrics to define success criteria for migration, as per the set application and organizational objectives. Therefore, it is useful in determining a suitable data fidelity level for a given migration scenario, as well as to confirm that the migration is successfully achieved as per the set quality objectives. These non-functional patterns are taken from related work \cite{Xu18, Wagner11} to form a complete pattern collection. However, they do not need extensive adaptations to use with blockchain data migration; hence, they are not presented in detail. In contrast, shaded boxes in the figure indicate the six new blockchain-specific data migration patterns we identified. We also present four other patterns from database migration, which need to be extended to work with unique properties of blockchains.

Multiple technical and process aspects such as consensus, immutability, state conversion, order of transactions, transaction fees, key management, and privacy need to be taken into account to preserve consistency, accountability, transparency, and provenance of data during the migration. Hence, the proposed patterns reflect a combination of data migration, process, and cloud architectural patterns. Moreover, these patterns could be applied to migrate data of a DApp and an entire blockchain alike. The intended audience for the proposed patterns is software architects, developers, system administrators, and project leads (both technical and business) who have to deal with the data migration life cycle from planning, design, execution to performance testing.

Next, we describe the patterns using the extended pattern form \cite{Meszaros97} and use the following eight elements to present essential details. The \textit{summary} presents a high-level description of what a pattern does. \textit{Context} defines the circumstances under which the problem exists and needs to be solved. \textit{Forces} are the contradictory considerations that must be taken into account while solving the problem under the given context. \textit{Problem} and \textit{solution} have the usual meaning where they specify what aspect of migration to be solved and how to solve it. \textit{Consequences} illustrate how the proposed solution resolves the forces while addressing the problem. Other patterns that may be of interest while solving the bigger problem are given under \textit{related patterns}. \textit{Known use} illustrates the applications of the pattern.

\subsection{State Extraction Pattern}\label{extract}
\subsubsection{Pattern 1 – Snapshotting}

\paragraph{\textit{Summary}.} Get a snapshot of states, smart contracts, and transactions on the source blockchain.

\paragraph{\textit{Context}.} The latest states and transactions of all accounts and smart contracts of the concert application need to be migrated to the target blockchain. Source blockchain is currently active; hence, the global state continues to change as new transactions arrive. The state to be migrated is already aggregated on the source blockchain.

\paragraph{\textit{Problem}.} How to get a complete account of states, smart contracts, and transactions on the source blockchain before migration?

\paragraph{\textit{Forces}.}
\begin{itemize}
	\item Anonymity – A blockchain does not track ownership of accounts, states, smart contracts, and transactions of an application due to anonymity.
	
	\item Consistency – Must capture the latest states and transactions of all accounts and smart contracts on the source blockchain. However, getting a globally consistent view of a distributed system is hard, e.g., it is difficult to freeze all transactions precisely at the same time on a distributed set of nodes.
	
	\item Finality – A state change may not be confirmed immediately after a transaction is included in a block.
	
	\item Latency – Time to collect states, smart contracts, and transactions are proportional to the number of accounts and smart contracts, as well as their states and transactions.
\end{itemize}

\begin{figure*}[htb!]
	\centering
	\includegraphics[width=\textwidth]{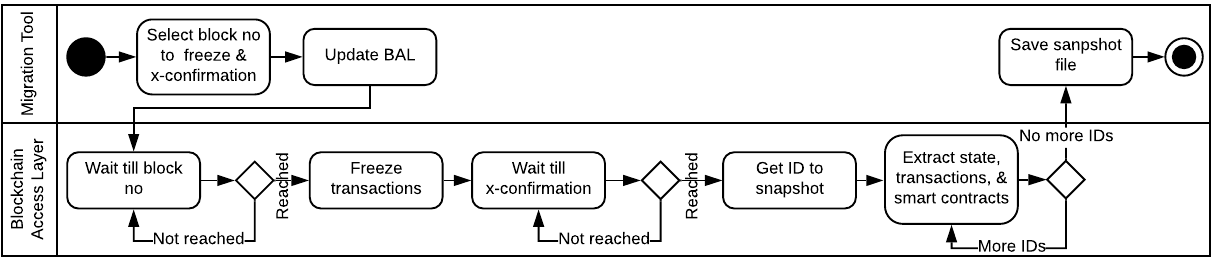}
	\caption{Process of taking a snapshot.\label{fig:snapshot}}
\end{figure*}

\paragraph{\textit{Solution}.} Get a snapshot of relevant states, smart contracts, and transactions on the source blockchain at a given time. Fig. \ref{fig:snapshot} shows the process of making a snapshot. First, select a block number to initiate the snapshotting process and number of blocks to wait for finality (aka., \texttt{x-confirmation}). Second, update all instances of the BAL. Third, the BAL should wait until the chosen block number is reached. Once reached, it should freeze processing further transactions to prevent any state changes. BAL should further wait for \texttt{x-confirmation} to ensure the finality of already issued transactions. Once it is reached, the migration tool can extract all required states, smart contracts, and  transactions by querying the source blockchain via the BAL. Finally, all extracted data are saved as a snapshot file.

Time is not reliable in blockchains due to clock skew. Therefore, time to freeze transactions and \texttt{x-confirmation} should be specified as block numbers because they are consistent across all blockchain nodes. If the entire blockchain is being migrated, all blockchain nodes need to freeze transaction processing at the set block number. In case the BAL or nodes are not designed to freeze transactions at a set number, they need to be patched (aka., soft fork). Such changes usually require approval from the blockchain governance body. Further considerations include the time needed to develop and patch all the BAL instances or nodes, as well as notify cryptocurrency exchanges and users about the potential downtime. Given such technical, operational, and governance considerations, the governance body should decide on the block number to freeze transactions well-ahead. For example, the typical time frame for a public blockchain migration range from weeks to months. \texttt{x-confirmation} is essential in Nakamoto consensus \cite{Nakamoto08} based blockchains, as the finality is probabilistic. A relatively higher value of \texttt{x-confirmation} is used to minimize the probability of state change after migration. For example, while Bitcoin and Ethereum typically use \texttt{x-confirmation}s of six and 12, respectively, they are almost doubled during migration. The time that the BAL or blockchain nodes remain frozen can be minimized by first taking a snapshot of the relevant data after the set block number and \texttt{x-confirmation}, then iteratively going through the new blocks that were generated since then to find only the updated state. If only the application-specific data are migrated, corresponding accounts can be found from the ID database. States represented as native assets of a blockchain platform, transactions, and smart contract code can be queried using the API exposed by the blockchain client nodes. If a smart contract exposes \texttt{getter} functions, its embedded state can be easily extracted. Otherwise, states have to be extracted by going through the global state data structure of a blockchain node. If the entire blockchain is to be migrated, go through the global state and entire block history to extract relevant accounts, states, smart contracts, and transactions. While a blockchain explorer can simplify this process, it is essential to validate that the explorer is in sync with the source blockchain. All data extractions can be parallelized as they are read-only.

\paragraph{\textit{Consequences}.}
\begin{itemize}
	
	\item Consistency is preserved as all blockchain activities are frozen on distributed instances of the BAL or blockchain nodes, and the snapshot is created after reaching time to finality.
	
	\item 
	Transactions could be made to freeze only while getting the snapshot of states, smart contracts, and transactions that were updated after initiating the snapshotting process. Thus, a snapshot on a blockchain can be made relatively faster than in a database, which needs to freeze all transactions due to the difficulty in determining the data that got changed once the migration process begins.
	
	\item ID database simplifies the identification of application-specific data overcoming anonymity.
	
	\item This pattern can be applied to states, smart contracts, and transactions, and any combination thereof.
	
	\item Latency depends on the time to finality and freeze time. Freeze time depends on the parallelization of state, smart contract, and transaction extraction. 
	Hence, this pattern is more desirable when the source blockchain is private, or the application does not interact with any external state.
\end{itemize}

\paragraph{\textit{Related patterns}.} To recreate the collected states, smart contracts, and transactions on the target blockchain use \textit{state and transaction load} patterns, such as \textit{establish genesis}, \textit{hard fork}, and \textit{state initialization}. Also, \textit{state transform}, as well as \textit{state and transaction load} patterns could use the snapshot to decide on what states and smart contracts to transform and load to the target blockchain.

\paragraph{\textit{Known use}.} Snapshotting is commonly used to bootstrap blockchains, sidechains, and new nodes. For example, a snapshot file format and bootstrap procedure for Bitcoin alt-coins is presented in \cite{Peter14}. EOS \cite{Milano18}, Telos \cite{Telos18}, TOMO \cite{Tomo18}, Tron \cite{Milano18}, and Vechain \cite{Vechain18} alt-coins took a snapshot while freezing all the transactions when they moved away from the Ethereum. Bithereum \cite{Bithereum18} took a snapshot of the Ethereum global state to spin up a duplicate blockchain (aka., hard spoon). Moreover, both Bitcoin and Ethereum support dumping the history at different granularities and bootstrapping a new node using the dumped snapshot. Such daily snapshots are available on the web. Snapshots are also used while migrating a DApp to a different smart contract or blockchain instance. For example, Tronbet took a snapshot of all ANTE tokens before exchanging them to the upgraded WIN tokens managed by a new smart contract \cite{Tronbet19}.

\subsection{State Transformation Patterns}\label{transform}
\subsubsection{Pattern 2 – State Aggregation}

\paragraph{\textit{Summary}.} Aggregate a set of states into a single (or a few) state.

\paragraph{\textit{Context}.} Concert application has a large number of accounts and states. Extracting all states from the source blockchain and recreating them on the target blockchain is both costly and time-consuming. 
The list of states to be migrated is given in the snapshot or ID database.

\paragraph{\textit{Problem}.} How to extract and recreate a large number of states while minimizing the time and cost?

\paragraph{\textit{Forces}.}
\begin{itemize}
	\item Size – A large number of accounts and their states to be migrated to the target blockchain.
	
	\item Latency – Time to extract and recreate the states is proportional to the number of accounts and their states.
	
	\item Cost – On a public blockchain, each state transfer needs to pay a transaction or exchange fee. The fee could also be proportional to the size of the state.
	
	\item Consistency – Must capture all accounts and states on the source blockchain. Any transformation of data before/after the migration must not violate consistency property.
	
	\item Accountability – Any data transformed to simplify the migration must be recorded with proof.
\end{itemize}

\begin{figure}[htb!]
	\centering
	\includegraphics[width=0.65\textwidth]{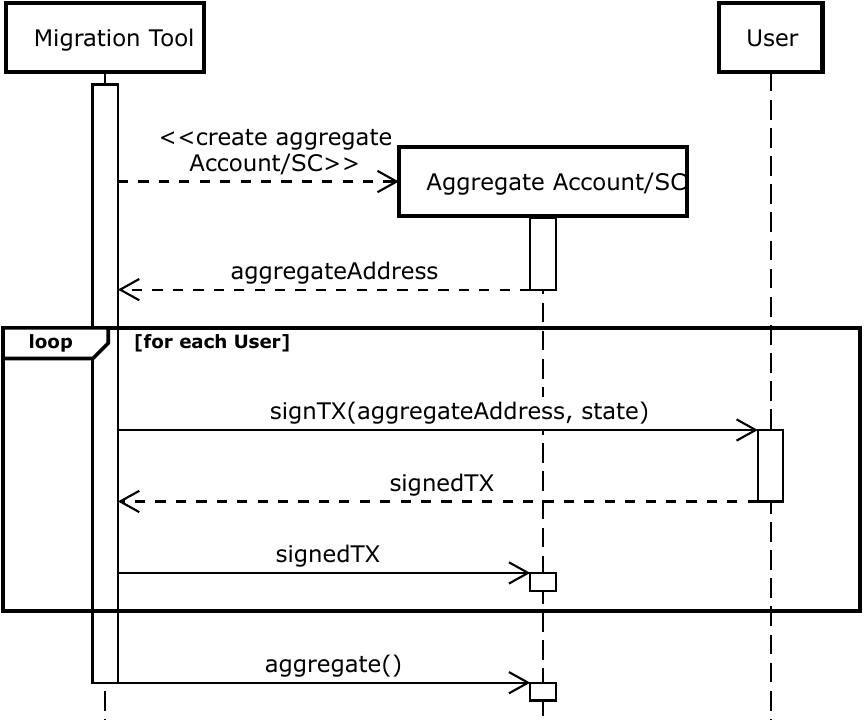}
	\caption{Object interaction during state aggregation.\label{fig:aggregation}}
\end{figure}

\paragraph{\textit{Solution}.} Transfer all ConcertCoins to a single account, such that only the closing balance needs to be migrated. Fig. \ref{fig:aggregation} shows the sequence of activities required to perform such an aggregation of state. To aggregate blockchain native assets, first, create a new account on the source blockchain. To aggregate states embedded in smart contracts, deploy a new Smart Contract (SC). Both cases will produce a new address (\texttt{aggregateAddress}). Then get the users' consent to transfer their states by signing a transaction with the current state as payload and \texttt{aggregateAddress} as the recipient. Next, submit the signed transaction to the source blockchain. Finally, trigger the \texttt{aggregate} function at the \texttt{aggregateAddress}.

The summation of native assets will be atomically performed once a transaction is executed. Hence, native asserts do not need a separate \texttt{aggregate} function. Smart contracts could be used to perform more generic aggregation scenarios. For example, a smart contract can aggregate fungible tokens, similar to native assets. Non-fungible tokens that could be mapped to a set of binary states can be concatenated to a bitmap represented as a single state. In a more general case, \textit{n} set of states could be aggregated to \textit{m} set of states ($m << n$), e.g., closing account balance vs. closing balances of each budget item. Once the \texttt{aggregate} function is called, it must lock the account or smart contract using the \textit{token burning} pattern to prevent further aggregations. When a user holds the private key, he/she needs to sign the transaction via the application or wallet software. It is pragmatic for the migration tool to submit transactions to the blockchain on behalf of the users, as it can track the transactions and trigger the \texttt{aggregate} function as soon as all the transactions are included. In practice, this is achieved by integrating the BAL or user wallet with the migration tool. Either the ID database or snapshot can be used to find the list of accounts and their states to be aggregated. Unless the application can work with the aggregated state(s) on the target blockchain (e.g., closing balance), the state needs to be disaggregated to match the states that were on the source blockchain. Disaggregation could be achieved by running relevant inverse functions and smart contracts on the target blockchain, e.g., ConcertCoins can be split into multiple accounts as per the state recorded in the snapshot. Aggregation or disaggregation could also be performed after extracting states from the source blockchain depending on the transaction fees, performance, and complexity of functions. Loss of accountability due to such external transformations can be overcome by adding the list of aggregations as a PoE entry to the blockchain. 
Transaction fees need to be paid in public blockchains to transfer states and run the \texttt{aggregate} function. While public blockchain users typically bear this cost, sometimes they are reimbursed while recreating the state on the target blockchain. Alternatively, the migration tool could top-up/reimburse user accounts before/after the state transfer \cite{Zhang19}.

\paragraph{\textit{Consequences}.} 
\begin{itemize}
	\item The number of accounts, their states, data size, and latency to extract and recreate are reduced.
	
	\item Consensus and the ability to record aggregation/disaggrega-tion operations within the blockchain provide greater consistency and accountability compared to database migration.
	
	\item In public blockchains, the cost could be reduced as transactions required for aggregation within the same blockchain is typically lower than the exchange fees or inter-blockchain transactions. However, the cost depends on the internal and external transaction fees, as well as the complexity and number of transactions and smart contracts required for aggregation/disaggregation.
	
	\item Transaction fees need to be either paid separately or could be deducted from the native assets being aggregated.
		
	\item While this pattern works on any blockchain, it works only on states that could be reduced to a single or a few values.
\end{itemize}

\paragraph{\textit{Related patterns}.} Aggregation can be performed before or after \textit{snapshotting} and before \textit{token burning}. Moreover, the set of states to be aggregated can be found using the \textit{snapshotting} pattern. Relevant \textit{state and transaction load} patterns could rely on this pattern to reduce the number of accounts and states to recreate. \textit{Off-chain data storage} pattern is needed to add a PoE entry when the aggregation is performed outside the source blockchain.

\paragraph{\textit{Known use}.} Aggregation is used while bootstrapping a token on the target blockchain. For example, Storj \cite{Storj17} and Binance \cite{Arron19} used the aggregated token balance from their source blockchains to set the initial balance of ERC-20 and BEP-2 token contracts, respectively. This pattern was also used in Ethereum DAO hard fork to aggregate blocked funds and transfer them back to the users \cite{Wilcke16}.  When the Bitcoin network is less congested, multiple UTXOs can be aggregated to produce a single UTXO by paying a small transaction fee. The resulting UTXO can be later spent during times of high network congestion while paying lower traction fee, as the overall transaction size is reduced \cite{Bitcoin18}.

\subsubsection{Pattern 3 – Token Burning}

\paragraph{\textit{Summary}.} Make states and smart contracts on the source blockchain unusable before the migration.

\paragraph{\textit{Context}.} Because the source blockchain used by the nonprofit is public, it is not decommissioned after the migration. Therefore, any state and smart contracts left in the source blockchain could be misused (e.g., double spending). 
The list of states and smart contracts to be migrated is given in the snapshot or ID database.

\paragraph{\textit{Problem}.} How to prevent the misuse of states and smart contracts after migration?

\paragraph{\textit{Forces}.}
\begin{itemize}
	\item Immutability – State of accounts and smart contracts on a blockchain is immutable.
	
	\item Consistency – If the source blockchain is not decommissioned, states and smart contracts could be used in both the blockchains leading to misuse, such as double-spending attacks \cite{SFOX19}. Any transformation of data to prevent misuse must not violate consistency properties.
	
	\item Accountability – Any data transformation to prevent misuse must be recorded with proof.
\end{itemize}

\paragraph{\textit{Solution}.} Use transactions to transfer states such as native assets and tokens to an unusable account. 
Delete tokens by calling respective functions on the smart contracts that created them. Similarly, smart contracts can call the self-destruct function. 
All such attempts to make states and smart contracts unusable are referred to as \textit{token burning}. 

While transferring states/assets, it is vital to ensure that the recipient address (aka., burn address) is invalid, i.e., there will not be a corresponding private key that can control the states sent to that address. Therefore, it is essential to use the burn address recommended by the chosen blockchain platform. Another alternative is to deploy a smart contract that immediately self-destructs as soon as it receives a transaction. However, it is more costly as smart contract deployment and execution are relatively expensive. A smart contract and its state can be made unusable using the self-destruct function, e.g., \texttt{selfdestruct} function in Ethereum. In case a self-destruct function is not built into the smart contract, a carefully crafted transaction(s) may be attempted to set the state to a terminating state \cite{Augur17}. However, this approach is not guaranteed to work or may fail in the future once vulnerabilities are found. Users holding the private keys need to sign the transactions that burn their states or trigger relevant smart contract functions. The burned state is typically verified before recreating the state on the target blockchain (aka., proof of burn) \cite{BitcoinWiki}. If a custodial wallet maintains aggregated state, a user trying to establish proof of burn for his/her state needs to transfer the state to a non-custodial wallet before burning the state. If the source and target blockchains are different instances, the service of an oracle could be used to attest the state from one blockchain to another. The migration tool may also work as an oracle, as it has access to both the blockchains. However, when transaction fees are charged, the burned state may not be an accurate representation of the original state. Sometimes token burning is performed after the migration, where they are retained as a rollback option in case of a failed migration. The risk of maintaining states on two blockchains needs to be carefully evaluated based on the permission type of blockchains, migration window, and the ability to prevent users from issuing transactions within the migration window. The risk could also be reduced by burning only one state at a time and recreating that on the target blockchain.

\paragraph{\textit{Consequences}.} 
\begin{itemize}
	\item Immutability, consistency, and accountability are preserved as all state changes are performed within the source blockchain. When recreating states on the target blockchain, proof of burn should be verified.
	
	\item This pattern works on all blockchains. On a public blockchain, transaction fees need to be paid to make the states and smart contracts unusable.
	
	\item Smart contracts should either implement a self-destruct function or it should be possible to generate a specific transaction sequence that sets a smart contract's state as terminated.
\end{itemize}

\paragraph{\textit{Related patterns}.} \textit{Snapshotting} and \textit{state aggregation} patterns could be used before \textit{token burning}. Moreover, the set of states and smart contracts to burn can be found from the snapshotting pattern. \textit{State and transaction load} patterns can verify proof of burn before initializing a new state.

\paragraph{\textit{Known use}.} While migrating from Ethereum or Bitcoin to their own instances, Binance \cite{Arron19}, Bitizens \cite{BitGuild18}, KARMA \cite{KARMA19}, Kin \cite{Kin18}, Safex \cite{Richard18}, and Storj \cite{Storj17} asked users to burn their tokens by transferring them to a designated address. Whereas EOS \cite{Milano18}, Qubicle \cite{Qubicles19}, and Vechain \cite{Vechain18} asked users to burn a small number of tokens to confirm the ownership of their accounts, and the remaining tokens were burned by the respective smart contracts that created them. Qubicle's token burning followed the EOS inter-blockchain communication protocol \cite{EOS18, Sigman}. While tokens are typically burned before the migration, Binance, KARMA, and Storj tokens were burned after the migration. This pattern is also used while bootstrapping alt-coins in Bitcoin \cite{BitcoinWiki} and sidechains in Horizen \cite{Garoffolo18}.

\subsection{State and Transaction Load Patterns}\label{load}
\subsubsection{Pattern 4 – Node Sync}

\paragraph{\textit{Summary}.} Create a clone of a blockchain node by synchronizing blocks and history.

\paragraph{\textit{Context}.} The nonprofit wants to add new nodes or swap nodes to change the physical location, improve performance, or create multiple instances of the source blockchain. A snapshot of the source blockchain is available.

\paragraph{\textit{Problem}.} How to clone a blockchain node?

\paragraph{\textit{Forces}.} 
\begin{itemize}
	\item Consistency – It is difficult to create a complete copy of a blockchain node, as the system is distributed, new blocks are built continuously, and finality is not immediate.
	
	\item Size – A blockchain may contain a large number of accounts, states, smart contracts, and transactions. Moreover, history may consist of a large number of blocks.
	
	\item Latency – Given the large size of the global state and history, time to make a copy could be very long. It takes even more time to recreate and validate the global state by applying all transactions and running smart contracts.
\end{itemize}

\begin{figure}[htb!]
	\centering
	\includegraphics[width=0.7\textwidth]{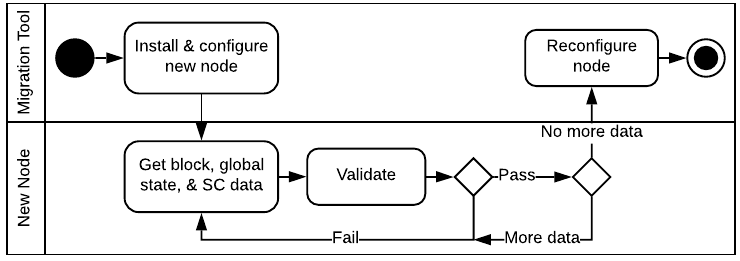}
	\caption{Process of synchronizing a node.\label{fig:sync}}
\end{figure}

\paragraph{\textit{Solution}.} Steps shown in Fig. \ref{fig:sync} can be used to synchronize the new node using the blockchain platform-specific sync tool. First, install the source blockchain platform's client software (or an updated version that is backward compatible) on the new node. Also, configure the new node to connect to other members of the source blockchain. Second, enable the sync tool on the node to copy various data structures representing the global state, smart contracts, transactions, and blocks from other blockchain nodes. Next, the node should rebuild and validate all the transactions from the genesis to verify the global state. 
Any errors, such as failed data transfer, need to be resolved by requesting further data. Finally, reconfigure the node to accept new transactions. 

The volume of the global state, smart contract, transaction, and block data of a highly-active blockchain could be in excess of hundreds of Gigabytes. Hence, many days to weeks could be required to download and validate them depending on bandwidth, IO, and CPU limitations, as well as ongoing transactions. Several optimizations could be applied to reduce the sync time. For example, the new node can be bootstrapped using a snapshot of the source blockchain. Then only the data from new blocks created after the snapshot need to be synchronized. Moreover, validation could be performed at different levels, such as validation of the last \textit{n} blocks, random blocks, entire block history, and entire block history and transactions. If the objective is to swap blockchain nodes, decommission the old nodes. Whereas if the objective is to create multiple blockchain instances, reconfigure the new node(s) to behave as an independent blockchain after a set block number. This is usually achieved by setting a new blockchain ID and disconnecting from the parent blockchain's nodes.

\paragraph{\textit{Consequences}.}
\begin{itemize}
	\item Blockchain sync tools rely on asynchronous messaging, immutability (to identify states updated after a given block number), consensus (to provide consistency), and validation. Hence, no state, history, and IDs are lost or changed in the process, ensuring consistency, integrity, and immutability.
	
	\item Time to synchronize a large volume of states and history could vary from hours to days. Time can be substantially reduced by initiating with a snapshot and synchronizing only the ledger state (i.e., \textit{state only} data fidelity level).
	
	\item This pattern works on any blockchain where source and target blockchain platforms are compatible, and the target node(s) is new.
\end{itemize}

\paragraph{\textit{Related patterns}.} \textit{Snapshotting} pattern can be used to speed up the bootstrap process.

\paragraph{\textit{Known use}.} This pattern is frequently used to connect new nodes to an existing blockchain. 
Bitcoin, Ethereum, and Bithereum \cite{Bithereum18} allow new nodes to download a snapshot file that reflects the global state and blocks at a set block number, and then sync with other nodes to speed up the process. Moreover, Ethereum supports syncing data at different granularity levels, such as full, fast, and light \cite{Buterin14}. Similarly, Bitcoin supports full node and thin client syncing.

\subsubsection{Pattern 5 – Establish Genesis}

\paragraph{\textit{Summary}.} Set the state on the target blockchain's genesis block.

\paragraph{\textit{Context}.} Concert application has a large number of states, and all of them need to be recreated on the target blockchain. The nonprofit has decided to spin up a new blockchain instance as the target and use ConcertCoin as the native asset. The list of accounts and states to be migrated is given in the snapshot. After the snapshot, states are marked as unusable using token burning. Relevant smart contracts are also redeployed.

\paragraph{\textit{Problem}.} How to load states to the target blockchain?

\paragraph{\textit{Forces}.}
\begin{itemize}
	
	\item Consistency – States such as blockchain native assets cannot be arbitrarily created because migration must preserve system invariants.
	
	\item Size – A large number of accounts and their states to be initiated on the target blockchain.
	
	\item Cost – Each account creation and state assignment needs to pay a transaction fee. The transaction fee could also be proportional to the size of the state.
	
	\item Latency – Time to create accounts and initialize their states is proportional to the number of accounts and their states.
	
	\item Accountability – Initiation of new accounts and their states on the target blockchain must be recorded with proof.
\end{itemize}

\begin{figure}[htb!]
	\centering
	\includegraphics[width=0.7\textwidth]{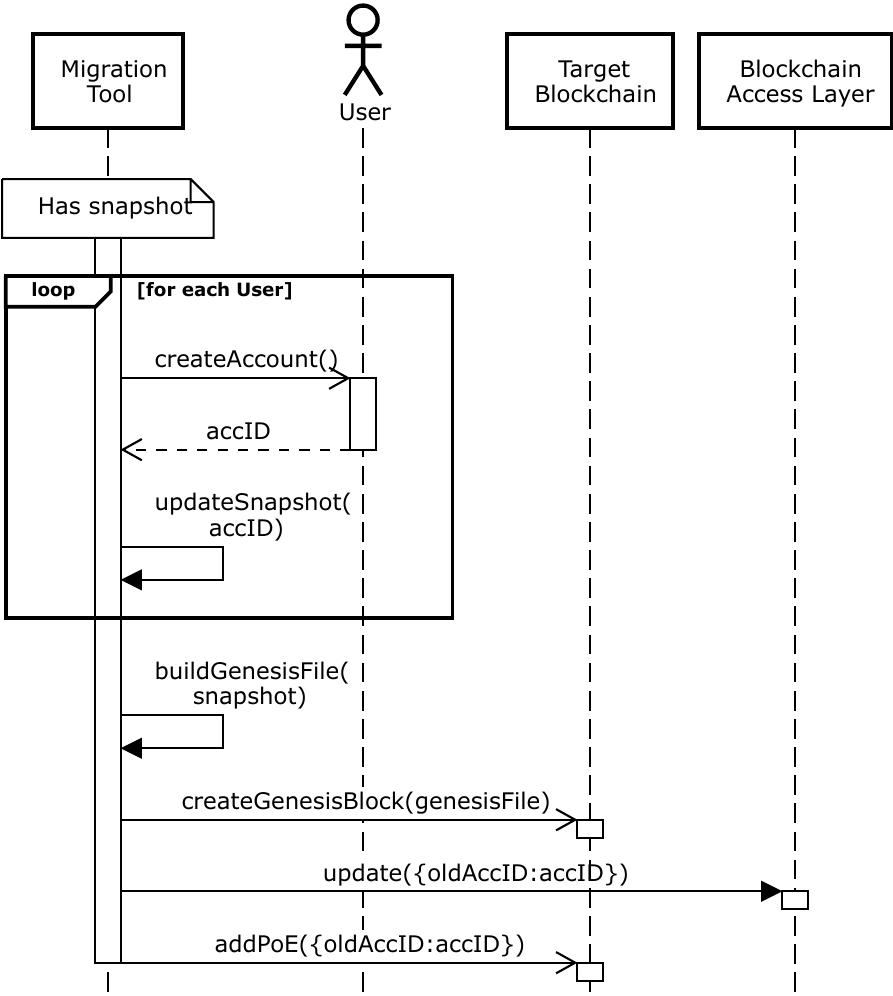}
	\caption{Object interaction to create the genesis block.\label{fig:genesis}}
\end{figure}

\paragraph{\textit{Solution}.} As seen in Fig. \ref{fig:genesis}, use the snapshot of states from the source blockchain to set states on the target blockchain's genesis block during its initialization. If a user cannot use an existing private key to prove its ownership of a state migrated to the target blockchain, a new key pair needs to be created. Thus, the first step is to get each user to create a new key pair and a corresponding account ID (\texttt{accID}) using the account creation algorithm of the target blockchain. Second, update the account ID in the snapshot file with \texttt{accID}. Next, create the genesis block configuration file (aka., genesis file) while including the accounts and states from the updated snapshot file. Then use the genesis file to initialize the target blockchain. Update the ID database on the BAL to reflect the new set of \texttt{accID}s. Also, add a PoE entry that tracks the mapping between old and new account IDs to record how the new accounts came into existence.

The first two and last two steps are optional when the accounts can be reused across the blockchains. However, it is desirable to create new private and public key pairs, as well as corresponding accounts on the target blockchain. This can guard against weaknesses that may appear in the source blockchain, its wallets, and exchanges that could compromise a private key or seed used to generate it. Further, the use of new accounts could enhance the anonymity of the application's accounts, especially when states are not very specific such that it is nontrivial to build a mapping table, e.g., when \textit{state aggregation} pattern is used.  User involvement is needed to create key pairs and new accounts if they hold the private keys. Moreover, platforms such as Hyperledger requires users to enroll their digital certificates with the certificate authority. If the number of states in the snapshot is too big to fit into the genesis block, \textit{aggregation} pattern could be used to reduce the state. Native assets are typically set during the genesis block creation. In practice, smart contracts are not deployed during the genesis. For example, even though Hyperledger Fabric \cite{Hyperledger19} does not use a native asset, smart contracts are not deployed using the genesis block. \textit{Encrypting on-chain data} pattern can be used to encrypt the PoE entry added to the blockchain to hide the mapping between old and new accounts. Time taken by all users to create new accounts is difficult to predict. Hence, sufficient notice ranging from a couple of days to weeks needs to be given to the users before the snapshot and genesis block are created.

\paragraph{\textit{Consequences}.} 	
\begin{itemize}
	
	\item Arbitrarily creating states on the genesis block to bootstrap a blockchain is not considered a consistency violation, as the genesis block is built by the blockchain governing body and the content of the block is transparent.
	
	\item Migration can be performed fast with no cost, as only the genesis block is used to recreate the state. However, the process slows down when users have to create new accounts.
	
	\item Accountability could be preserved by adding the PoE entry that tracks the mapping between old and new account IDs. 
	
	\item This pattern can be used to recreate native assets of any new instance of a blockchain.
	
	\item The pattern works only if the states fit into a single block or can be aggregated to fit into a single block.
\end{itemize}

\paragraph{\textit{Related patterns}.} \textit{State extraction} and \textit{state transformation} patterns could be used to identify, reduce, and burn the states to include in the genesis block. Use the \textit{off-chain data storage} and \textit{encrypting on-chain data} patterns to add an encrypted PoE entry that tracks the mapping between old and new account IDs.

\paragraph{\textit{Known use}.} The genesis block is used to specify the initial distribution of native assets on a blockchain. For example, the genesis block of the Æternity \cite{Aeternity19} included the ERC-20 tokens of users who supported the initial migration from Ethereum to its blockchain. Zeepin \cite{Zeepin18} used a similar approach as it migrated from NEO to its blockchain. Telos \cite{Telos18} -- an instance of the EOS platform -- used the EOS snapshot file to build its genesis block. The pattern is also used for hard spooning, e.g., Bithereum \cite{Bithereum18} created its genesis block using a snapshot from Bitcoin at a set block number.

\subsubsection{Pattern 6 – Hard Fork}
\paragraph{\textit{Summary}.} Change the global state of the target blockchain.

\paragraph{\textit{Context}.} Concert application has a large number of states, and all of them need to be recreated on the target blockchain. The nonprofit has chosen an existing private or consortium blockchain as the target. The list of accounts and states to be migrated is given in the snapshot. After the snapshot, states are marked as unusable using token burning.

\paragraph{\textit{Problem}.} Same as pattern five.

\paragraph{\textit{Forces}.} Same as pattern five.

\begin{figure}[htb!]
	\centering
	\includegraphics[width=0.75\textwidth]{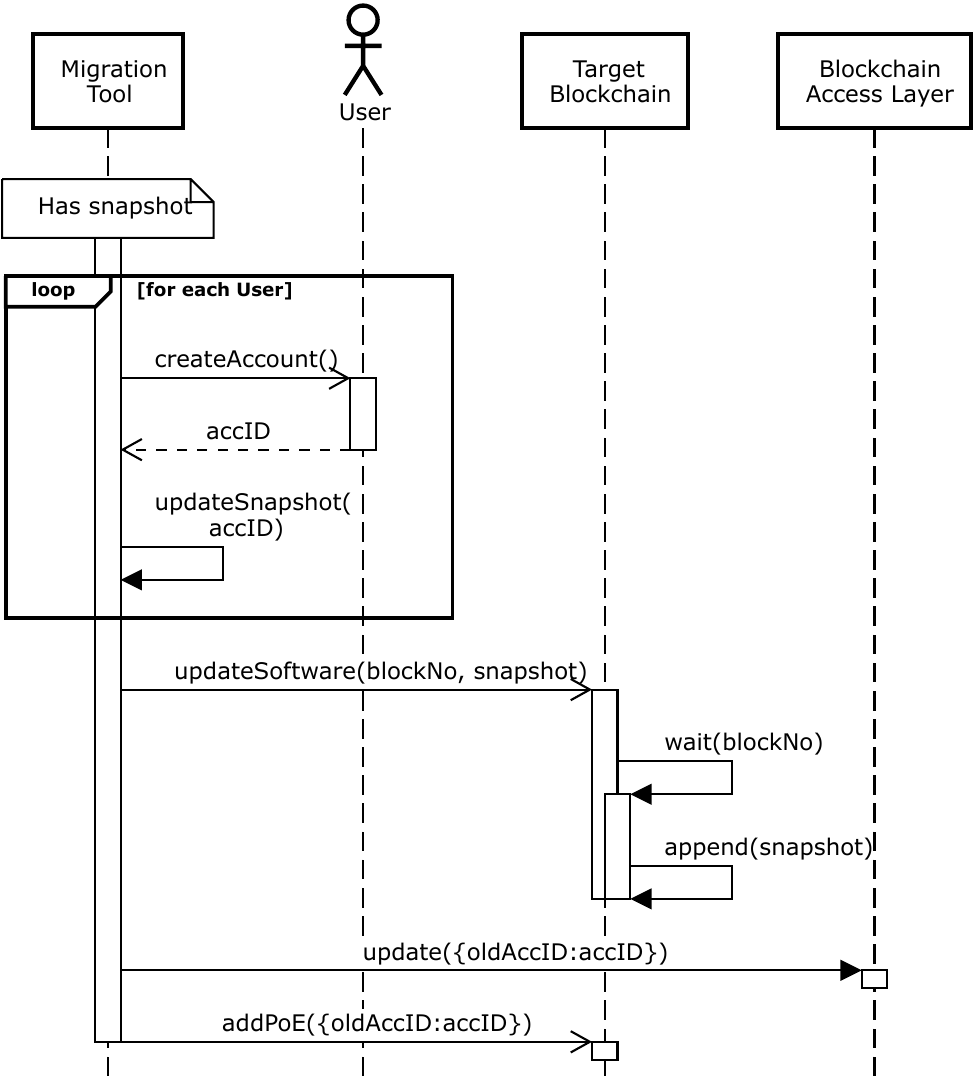}
	\caption{Object interaction during a hard fork.\label{fig:hardfork}}
\end{figure}

\paragraph{\textit{Solution}.} As seen in Fig. \ref{fig:hardfork}, use the states from the snapshot to change the global state on the target blockchain. If a user cannot use an existing private key to access the state migrated to the target blockchain, follow the first two steps to create a new account and update the snapshot file similar to the \textit{establish genesis} pattern. Next, update the blockchain client software on all nodes of the target blockchain to include the snapshot file at a set block number (\texttt{blockNo}). Once the \texttt{blockNo} is reached, all blockchain nodes should append the states from the snapshot file into the ledger to update the global state. Due to the arbitrary addition of new states, blocks produced before and after the software update will be incompatible. Such a change in blockchain state transition rules is referred to as a \textit{hard fork} \cite{Wilcke16}. Similar to the \textit{establish genesis} pattern, update the ID database and add a PoE entry to reflect new account IDs.

While the first two and last two steps are optional when the accounts can be reused across the blockchains, it is desirable to create new key pairs and accounts as discussed in the \textit{establish genesis} pattern. Similarly, user involvement is needed to create new accounts. While any state can be amended during a hard fork, deployment of smart contracts is unprecedented. However, compared to the \textit{establish genesis} pattern, there is no limit on the number of states that can be recreated during a hard fork. Similar to the \textit{snapshotting} pattern, approval of the blockchain governance body is needed to update the blockchain client software. Further, the \texttt{blockNo} to initiate the hard fork need to be determined while considering the time needed to develop and patch all the nodes, as well as time taken by all users to create new accounts. Hence, a notice period ranging from a couple of days to weeks needs to be given before the hard fork \cite{Hive20}.

\paragraph{\textit{Consequences}.} 
\begin{itemize}
	\item Migration happens at a set block number with no cost. Any node that misses the update will no longer be part of the same blockchain.
	
	\item Accountability could be preserved by adding the PoE entry that tracks the mapping between old and new account IDs.
	
	\item While this pattern works with any state, it violates consistency. Hence, the consensus of the blockchain's governance body is required to update the blockchain software to initiate the hard fork \cite{Wilcke16}. Therefore, it is more suitable for private and consortium blockchains.
	
\end{itemize}

\paragraph{\textit{Related patterns}.} Same as pattern five.

\paragraph{\textit{Known use}.} Hard forks are used to spin-up new cryptocurrencies. For example, after launching the blockchain using \textit{establish genesis} pattern, Æternity went through three rounds of hard forks to migrate the remaining ERC-20 tokens \cite{Aeternity19}. When Steem was sold to Tron, objecting users spin up a new cryptocurrency called Hive by forking the Steem blockchain \cite{Hive20}. Similarly, a conflict over the use of a hard fork to recover cryptocurrency lost due to the DOA attack \cite{Wilcke16} resulted in Ethereum and Ethereum Classic.

\subsubsection{Pattern 7 – State Initialization}

\paragraph{\textit{Summary}.} Initialize/recreate states on the target blockchain.

\paragraph{\textit{Context}.} Concert application has a large number of states and smart contracts, and all of them need to be recreated on the target blockchain. The nonprofit has chosen an existing public blockchain as the target. The list of accounts, smart contracts, and states to be migrated is given in the snapshot or ID database. After the snapshot, states are marked as unusable using token burning.

\paragraph{\textit{Problem}.} Same as pattern five.

\paragraph{\textit{Forces}.} Same as pattern five.

\begin{figure}[htb!]
	\centering
	\includegraphics[width=0.75\textwidth]{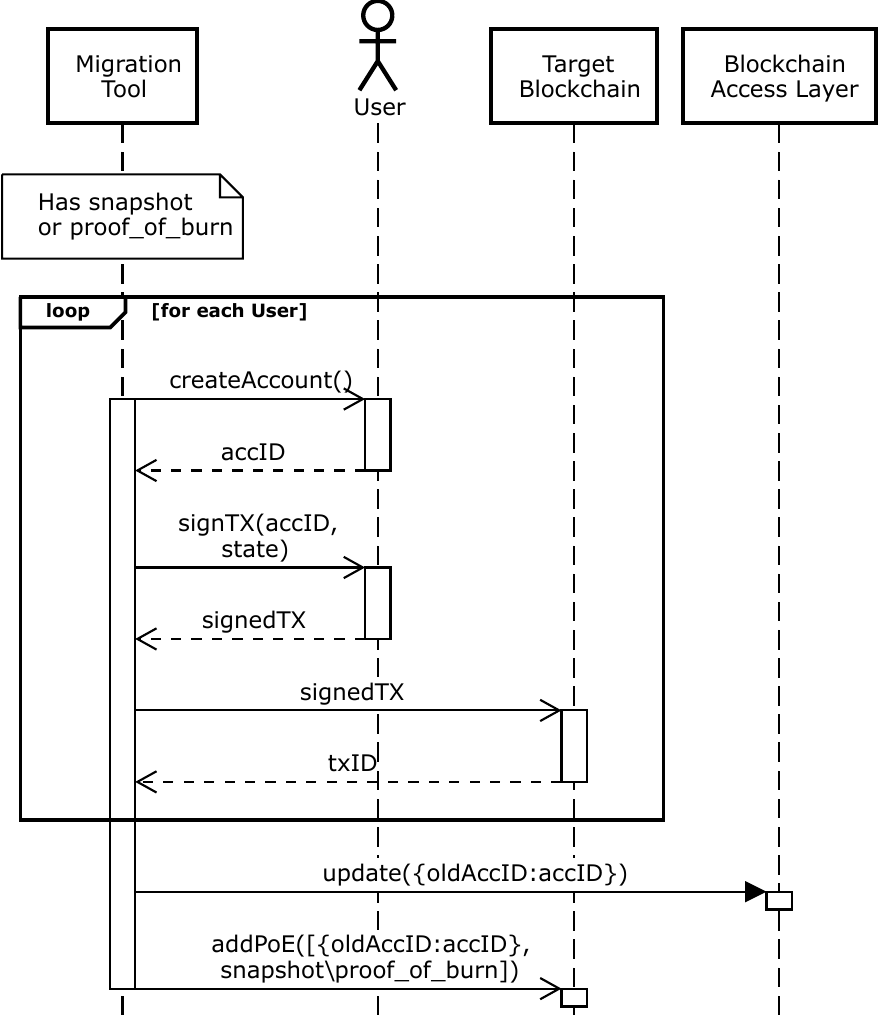}
	\caption{Object interaction during state initialization.\label{fig:state_init}}
\end{figure}

\paragraph{\textit{Solution}.} Create one state at a time on the target blockchain using the steps shown in Fig. \ref{fig:state_init}. If a user cannot use an existing private key to access the state migrated to the target blockchain, follow the first step to create a new account similar to the \textit{establish genesis} pattern. Then get the user to sign a transaction with the state listed in the snapshot as the payload and new address (i.e., \texttt{accID}) on the target blockchain as the recipient. Then submit the signed transaction to the target blockchain to recreate the state. Similar to the \textit{establish genesis} pattern, update the ID database, and add a PoE entry to reflect new account IDs. It is also desirable to include the snapshot file as a PoE entry, as it is not included in the target blockchain's history compared to \textit{establish genesis} and \textit{hard fork} patterns.

While the first and last two steps are optional when the accounts can be reused across the blockchains, it is desirable to create new key pairs and accounts, as discussed in the \textit{establish genesis} pattern. Also, user involvement is needed to create new accounts. When users hold the private keys, they also need to sign the transaction used to recreate the state. Signed transactions are issued by the migration tool to the target blockchain, mainly for the convince of managing the migration. For example, when the transactions flow through the migration tool, it is easier to update the ID database and use the \textit{measure migration quality} pattern to quantify the progress of migration. Instead, users may send the signed transactions directly, as each state is initiated using an independent transaction. In fact, the entire loop can be parallelized. Moreover, rather than using a snapshot file to identify the list of accounts and states to recreate, ID database could be used. However, it is essential that the latest state on the source blockchain is captured and recreated on the target blockchain. Thus, the \textit{token burning} pattern can be used to grantee that a state will not further change on the source blockchain. This pattern is suitable to redeploy smart contracts and recreate states embedded in them, e.g., tokens. Sometimes a series of transactions may need to be replayed to set a particular state. However, this is time-consuming and costly, as transactions need to be replayed in exact order with finality. However, native assets cannot be recreated due to the violation of systems invariants. Assets lost as transaction fees during \textit{state transformation}, \textit{token burning}, and setting state could also be credited back while initiating a state. Time taken by all users to create new accounts and sign transactions is difficult to predict. Hence, sufficient notice ranging from a couple of days to weeks needs to be given to the users \cite{Richard18, Vechain18, Milano18, Effect.AI19}. However, user involvement does not slow down the entire migration process, as each state is independently recreated.

\paragraph{\textit{Consequences}.}
\begin{itemize}
	\item One state can be initialized at a time without requiring the migration to be single-shot operation; hence, the risk is reduced.
	
	\item Consistency and accountability are preserved when proof of burn is verified before the state initialization and a PoE entry of the snapshot is added to the blockchain.
	
	\item This pattern works with any target blockchain regardless of its existence. However, it works only for smart contracts and states that can be arbitrarily created, e.g., tokens generated by smart contracts. It does not work on native assets.
	
	\item If the number of accounts and their states is large, many transactions are needed to set their initial state, increasing the cost and latency. Cost and latency could be reduced using the \textit{state aggregation} pattern.
\end{itemize}

\paragraph{\textit{Related patterns}.} Same as pattern five.

\paragraph{\textit{Known use}.} Binance \cite{Arron19}, Bithereum \cite{Bithereum18}, Bitizens \cite{BitGuild18}, Effect.AI \cite{Effect.AI19}, EOS \cite{Milano18}, Gifto \cite{Gifto19}, Kin \cite{Kin18}, Qubicle \cite{Qubicles19}, Safex \cite{Richard18}, Storj \cite{Storj17}, and Vechain \cite{Vechain18} migration used this pattern to issue new tokens on the target blockchain. Bithereum, Effect.AI, EOS, Kin, Qubicle, Safex, Storj, and Vechain also relied on users to create new accounts on the target blockchain, and inform new addresses to the migration tool. Other than Bithereum and Gifto, others also used \textit{token burning} in combination with this pattern.

\subsubsection{Pattern 8 – Exchange Transfer}
\paragraph{\textit{Summary}.} Transfer states via an exchange.

\paragraph{\textit{Context}.} Concert application has a large number of states, and all of them need to be recreated on the target blockchain. The nonprofit has chosen an existing public blockchain as the target and ConcertCoins are to be converted to its native assets. The list of states to be migrated can be found from the ID database.

\paragraph{\textit{Problem}.} Same as pattern five.

\paragraph{\textit{Forces}.} Same as pattern five.

\begin{figure}[htb!]
	\centering
	\includegraphics[width=0.75\textwidth]{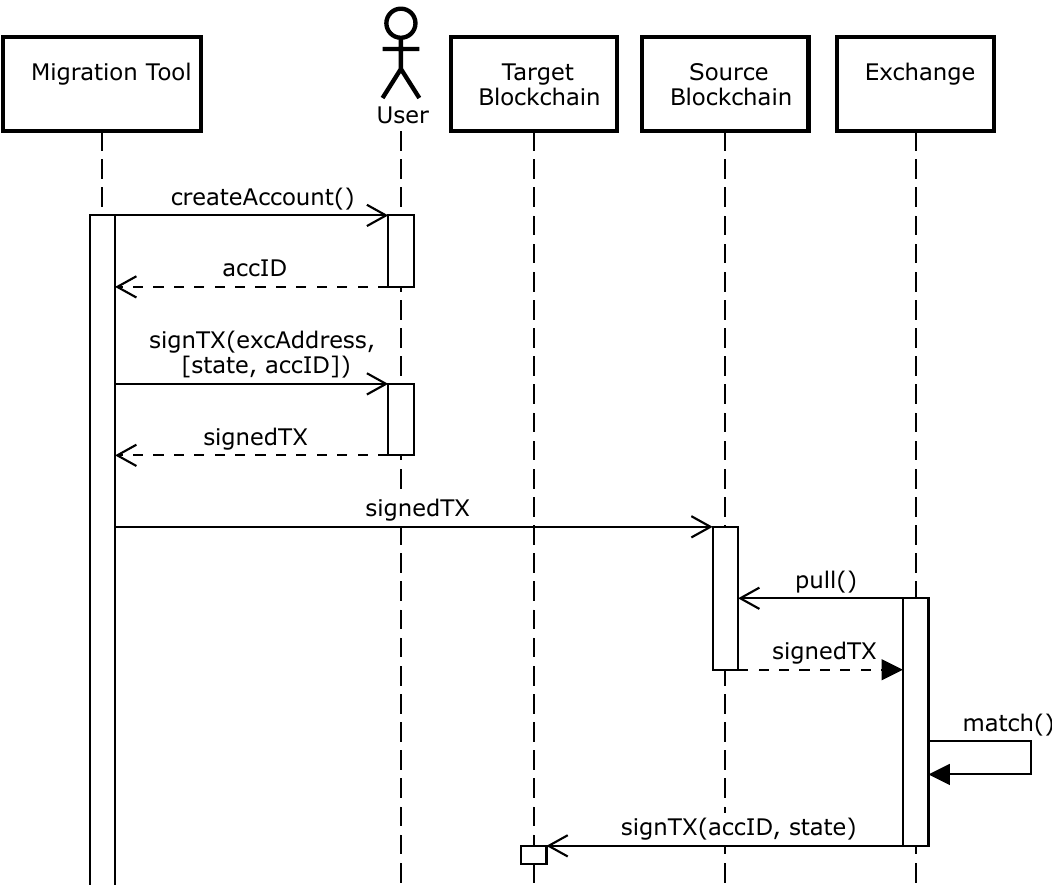}
	\caption{Object interaction during exchange transfer.\label{fig:exchange}}
\end{figure}

\paragraph{\textit{Solution}.} As seen in Fig. \ref{fig:exchange}, use a cryptocurrency/token exchange to transfer the states. If a user cannot use an existing private key to access the state migrated to the target blockchain, follow the first step to create a new account similar to the \textit{establish genesis} pattern. Then get the user to sign a transaction with both the state and new account ID (i.e., \texttt{accID}) on the target blockchain as the payload. Set the address of the exchange (\texttt{excAddress}) as the recipient. Then submit the signed transaction to transfer the state to exchange's account on the source blockchain. This is similar to a sell order in financial markets. Once the exchange confirms that the state is transferred to its account, it will look for a matching buy order. Once a match is made, the exchange will transfer the state to the \texttt{accID} using its account in the target blockchain. Similar to the \textit{establish genesis} pattern, finally update the ID database to reflect new account IDs (this step is not shown in Fig. \ref{fig:exchange}).

Some of the exchange-based data migration scenarios use a combination of \textit{token burning} and \textit{state initialization} patterns. Whereas Fig. \ref{fig:exchange} captures the behavior of an exchange matching sell and buy orders from source and target blockchains, which is more prevalent in decentralized exchanges. Chosen exchange(s) should support the state to be traded. If the format of states used by the application and exchange does not match, the state could be tokenized to a common format (e.g., ERC-20 standards in Ethereum). This could be achieved by deploying a smart contract to convert tokens on the source blockchain. Similarly, tokens could be decoded on the target blockchain. The number of public blockchains and types of assets/tokens supported by a centralized exchange are usually limited. However, a decentralized exchange can be configured to connect with any blockchain and support any token format agreeable to both the blockchains. Even the migration tool could be developed to act as a decentralized exchange. The first and last steps are optional when the accounts can be reused across the blockchains. However, it is desirable to create new accounts, as discussed in the \textit{establish genesis} pattern. Also, user involvement is needed to create new accounts. Similar to the \textit{state initialization} pattern, signed transactions are issued by the migration tool to the source blockchain, mainly for the convince of managing the migration. Instead, users may send the signed transactions directly, as each state is exchanged using an independent transaction. Moreover, as the centralized exchanges usually keep custody of a user's private key, it could automatically transfer the state without user involvement. In all scenarios, a transaction fee and commission are usually charged. Time taken by all users to create new accounts and sign transactions is difficult to predict. Similarly, the time needed by the \texttt{match} function depends on market demand for asset/state. Hence, sufficient notice ranging from a couple of weeks to months needs to be given to the users \cite{Vechain18}. However, this user involvement does not slow down the entire migration process, as each state is independently transferred.

\paragraph{\textit{Consequences}.}	
\begin{itemize}
	\item One state can be transferred at a time without requiring the migration to be single-shot operation; hence, the risk is reduced. Cost is proportional to the number of states and their sizes/values to transfer. Bulk/aggregated transfers and the use of centralized exchanges could reduce cost and latency.
	
	\item Because the state transfers to and from the exchange’s account are recorded on the blockchain, better consistency and accountability can be achieved. Decentralized exchanges provide enhanced privacy compared to centralized ones due to disintermediation.
	
	\item This pattern works with both blockchain native assets and tradable tokens on any blockchain platform. However, the state should be tradeable on the chosen exchange(s) or should be convertible to a tradable token.
	
	\item Exchanges charge transaction fees and apply asset conversion rates reducing the value of assets during the transfer.
\end{itemize}

\paragraph{\textit{Related patterns}.} Same as pattern five.

\paragraph{\textit{Known use}.} During the migration from Ethereum, Atomic \cite{Wright20}, Binance \cite{Arron19}, Gifto \cite{Gifto19}, TOMO \cite{Tomo18}, Tron \cite{Milano18}, and VeChain \cite{Vechain18} requested users to migrate their assets to a selected centralized exchange. ChangeNOW \cite{Wright20} token swap service was used by Atomic to migrate from Ethereum to Binance. KyberSwap also offers a similar service for ERC-20 tokens \cite{Kyber19}. EOS \cite{Sigman} and Interledger \cite{Thomas15} provide protocols to transfer tokens via decentralized exchanges.

\subsubsection{Pattern 9 – Transaction Replay}
\paragraph{\textit{Summary}.} Replay transactions on the target blockchain to recreate both states and history.

\paragraph{\textit{Context}.} Concert application has a large number of accounts, states, transactions, and smart contracts. All of which need to be recreated on the target blockchain while preserving the history of state changes. The initial state is already recreated on the target blockchain. List of transactions to be replayed can be found from the snapshot.

\paragraph{\textit{Problem}.} How to recreate both the states and history on the target blockchain?

\paragraph{\textit{Forces}.}
	\begin{itemize}
		\item Consistency – The order of transactions replayed on the target blockchain must be the same as the source blockchain. Moreover, the application may have sent/received transactions to/from accounts that are external to the application, e.g., users buying concert tickets.

		\item Volume – A large number of transactions need to be executed on the target blockchain.

		\item Latency – Time to execute transactions is proportional to the number of transactions and complexity of smart contracts.

		\item Cost – On public blockchains, transaction fees need to be paid. Fees could also be proportional to the size of states and complexity of smart contracts executed.
	\end{itemize}

\begin{figure}[htb!]
	\centering
	\includegraphics[width=0.7\textwidth]{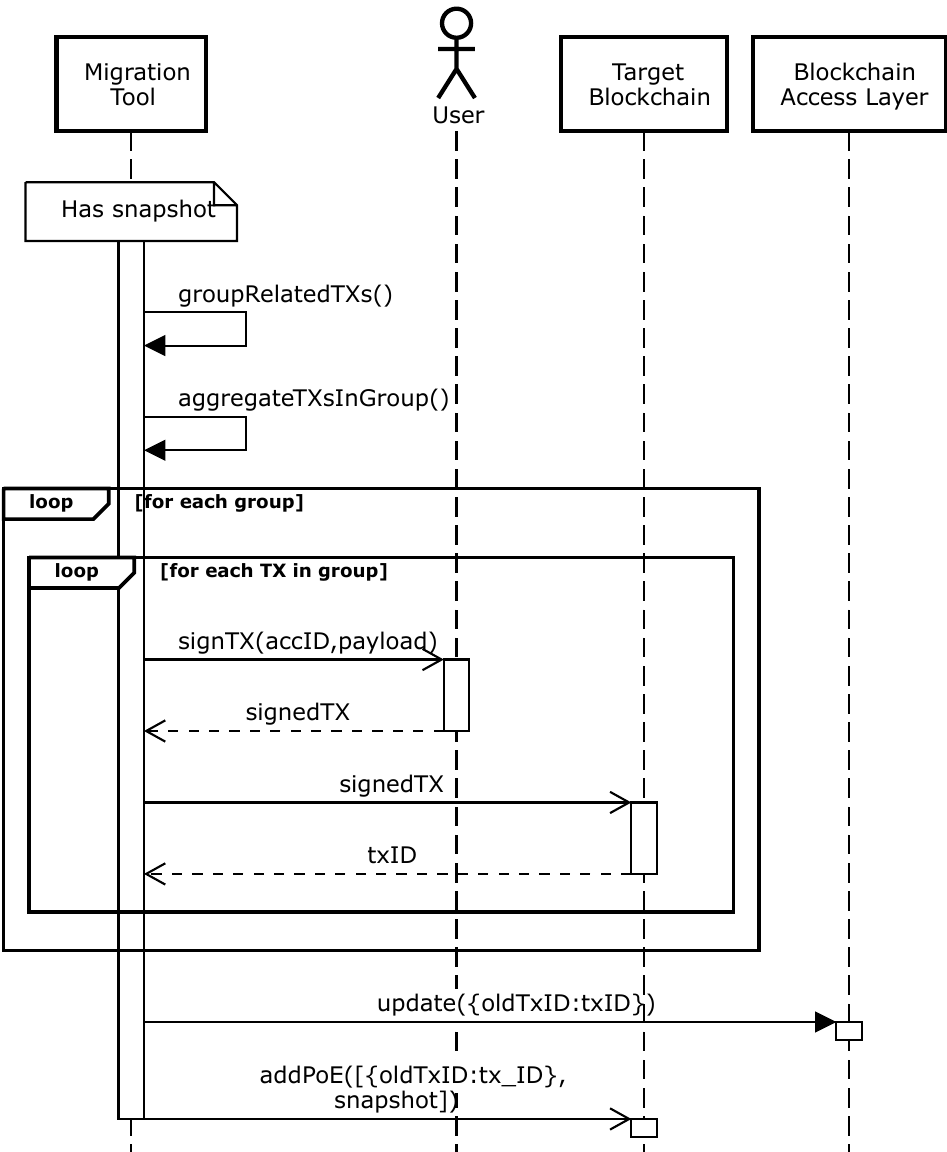}
	\caption{Object interaction during transaction replay.\label{fig:tx_replay}}
\end{figure}

\paragraph{\textit{Solution}.} Replay transactions from the source blockchain using the steps outlined in Fig. \ref{fig:tx_replay}. When replaying transactions on the target blockchain, their causal relationships must be preserved, e.g., the order of transactions send/received by an account may need to be preserved to maintain consistency of a state. Therefore, first group related transactions in the snapshot file. If abstraction is acceptable, also aggregate the transactions in a group such that the final state is consistent. For example, e.g., payments from $A \rightarrow B$ and $B \rightarrow C$ can be abstracted as $A \rightarrow C$. Third, concurrently replay each independent group of transactions on the target blockchain, while sequentially replaying all transactions within a group to preserve causal relationships. When the private keys or signature format of the source and target blockchains are different, transactions need to be signed again before replaying on the target blockchain. Address of the recipient also needs to be updated when the same addresses cannot be used across the blockchains. Then submit the signed transaction to the target blockchain. The transaction ID (\texttt{txID}) is likely to vary across blockchain instances, and also depends on nonce, sender and receiver addresses, and transaction fees. Hence, use the second last step to update the mapping between old and new \texttt{txID}s on the ID database on the BAL. Finally, add a PoE entry to reflect new transactions IDs. If translations are aggregated, also include the snapshot file as a PoE entry, as it is not included in the target blockchain’s history.

While the second step is optional, it can substantially reduce the time to replay transactions and associated transaction fees. Signing transactions, updating transaction IDs, and PoE entry are needed when the account and smart contract addresses are different across the two blockchains. Even if the addresses can be reused, it is desirable to create new addresses as discussed in the \textit{establish genesis pattern}. When users hold private keys, they need to sign the transaction. Compared to \textit{state initialization} and \textit{exchange transfer} patterns, it is essential that the migration tool issues transactions to the target blockchain on behalf of the users. For example, the order of transactions needs to be maintained. Moreover, if a transaction fails, it needs to be retried, and a dependent transaction can be replayed only when the finality of the previous transaction is confirmed. This is particularly required in blockchains such as Bitcoin and Hyperledger Fabric that do not use a sequence number to order transactions. Blockchains that uses a sequence number, e.g., nonce in Ethereum, can simultaneously issue an ordered set of transactions while substantially reducing time to replay transactions. Furthermore, the migration tool has better visibility to aggregate transactions from the snapshot. It is not possible to replay transactions issued by source addresses that are outside the application's control, e.g., a user that bought a concert ticket. Rather than fabricating a transaction, it is desirable to set such state using other \textit{state load} pattern that requires the approval of the blockchain governance body. Time taken by all users to sign transactions and time to replay a series of transactions is difficult to predict. Hence, sufficient notice ranging from a couple of days to weeks need to be given to the users. However, this does not slow down the entire migration process, as each related group of transactions is independently replayed.

\paragraph{\textit{Consequences}.}
	\begin{itemize}	
		\item Consistency is preserved as all transactions are replayed while maintaining their dependencies.

		\item This pattern works with any target blockchain regardless of whether it is new, pre-exists, or compatible.

		\item Latency depends on the volume of transactions, target blockchain's throughput, time to finality, concurrent transaction limit, and interdependencies among transactions.

		\item This pattern is suitable to recreate states embedded in a smart contract, e.g., tokens. However, native assets cannot be recreated as it violates systems invariants.

		\item This could be cost-prohibitive on a public blockchain, as fees are proportional to the number of transactions and data size. Fees could be reduced by aggregating transactions.
	\end{itemize}

\paragraph{\textit{Related patterns}.} Accounts, initial states, and smart contracts need to be created on the target blockchain using relevant \textit{state load} and \textit{smart contract} patterns before replaying the transactions. \textit{Snapshotting} pattern is required to capture the list of transactions to replay. \textit{Off-chain data storage} and \textit{encrypting on-chain data} patterns are needed to add a PoE entry of aggregated transactions and updates to the ID database.

\paragraph{\textit{Known use}.} In Ethereum and Hyperledger transaction replay is used to validate a blockchain node after synchronizing and rebooting. After Bitcoin and Ethereum hard forks that spit each platform into two instances, transaction replay has been used as an attack to double spend and change the state on both instances \cite{SFOX19}.

\subsection{Smart Contract Patterns}\label{sm}
\subsubsection{Pattern 10 – Virtual Machine Emulation}
\paragraph{\textit{Summary}.} Allow smart contracts written in one language to run on another blockchain platform.

\paragraph{\textit{Context}.} Concert application uses a set of smart contracts and embedded states which need to be usable on the target blockchain. The target blockchain platform could run/emulate the Virtual Machine (VM) used to execute smart contracts. The list of smart contracts and their states to be migrated is given in the snapshot. After the snapshot, smart contracts are marked as unusable using token burning.

\paragraph{\textit{Problem}.} How to run smart contracts written for one blockchain platform on another?

\paragraph{\textit{Forces}.}	
\begin{itemize}
	\item Platform dependence – Usually, the smart contract language is blockchain platform-specific.
	
	\item Turing completeness – Not all smart contract languages are Turing complete. Thus, it may not be possible to recreate the same behavior using a different smart contract language.
	
	\item Correctness – It is difficult to guarantee that the rewritten smart contract behaves precisely like the original smart contract.
	
	\item Time and cost – Rewriting and testing smart contracts take time hence also costly.
\end{itemize}

\begin{figure*}[htb!]
	\centering
	\includegraphics[width=\textwidth]{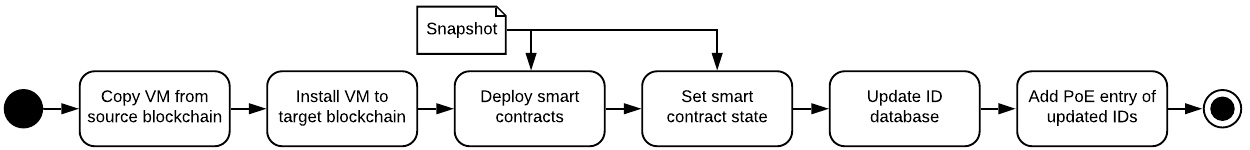}
	\caption{Process of virtual machine emulation.\label{fig:vm_emulation}}
\end{figure*}

\paragraph{\textit{Solution}.} Use the process outlined in Fig. \ref{fig:vm_emulation} to reuse the smart contract execution environment (aka., VM, sandbox, or container) on the target blockchain. First, copy the VM from the source blockchain. Second, integrate the VM into the target blockchain. Third, if the VM does not hold the smart contract code, redeploy the smart contracts on the target blockchain using the \textit{state initialization} pattern. Fourth, use the same pattern to set the states of the deployed smart contracts. Smart contract code and their states can be found from the snapshot. Sixth, update the mapping between old and new smart contract addresses on the ID database, as smart contract addresses vary across blockchain instances, and could also depend on the address that deployed the smart contract, transaction sequence number, among others. Finally, it is also desirable to include the snapshot file and mapping between old and new smart contract addresses as a PoE entry, as they are not included in the target blockchain's history.

If the target blockchain allows importing the source blockchain's VM, it could be copied over. However, in practice, it is more likely for the target blockchain to use a customized VM that supports the same instruction set. For example, while Hyperledger Burrow supports the Ethereum VM (EVM), a proxy is used to abstract the transaction fee-related parameters, as Hyperledger does not have a concept of transaction fees. Moreover, not all VMs (e.g., EVM) hold the smart contract code. While Hyperledger chaincode containers hold the code, a smart contract still needs to be instantiated on the target blockchain before use. Such instantiation may also be used to set the address and access rights of a smart contract. Moreover, even if the addresses can be reused, it is desirable to create new accounts, as discussed in the \textit{establish genesis pattern}. This intern could change the address of the smart contract. Therefore, rather than copying the VM, which is likely to require extensive configuration, it is easier to redeploy the smart contracts on the target blockchain's VM. In such cases, the first two steps can be skipped, given that the two VMs exhibit the same execution behavior. Such compatibilities could be checked using bytecode-level formal verification while reducing the cost and complexity of smart contract validation \cite{Amani18}. When account owners hold the private keys, users need to create new accounts, redeploy smart contracts, and set their states. Also, if the reference to a smart contract is provided via a smart contract registry \cite{Xu18}, first, the registry contract needs to be deployed. Then update the registry with the new smart contract address.

\paragraph{\textit{Consequences}.} 	
\begin{itemize}	
	\item Execution behavior and correctness of the original smart contract are preserved, as the same smart contract code and execution environment are used.
	
	\item Saves time and reduces cost, as no code reverse engineering, translation, and testing are needed.
	
	\item Works only when the target blockchain platform can emulate source blockchain's VM (i.e., smart contract platform independence) without any limitations in behavior and Turing completeness.
\end{itemize}

\paragraph{\textit{Related patterns}.} When emulation is not possible, use the \textit{smart contract translation} pattern. \textit{State extraction} and \textit{state transformation} patterns can be used to capture the smart contract code, embedded states, aggregate the states, and burn the smart contracts. Use the \textit{state initialization} pattern to deploy and set the state of smart contracts. PoE entries of new smart contract addresses can be added using the \textit{off-chain data storage} and \textit{encrypting on-chain data} patterns.

\paragraph{\textit{Known use}.} EVM bytecode can run on Hyperledger \cite{Hyperledger18}, VMware Concord \cite{Gueta19}, Tron, and VeChain. For example, Deloitte was able to move smart contracts from Ethereum to VeChain, as both blockchains supported the same EVM \cite{Leonowicz19}. Docker containers used by Hyperledger to store and execute smart contract code could be reused on multiple channels and across different networks.

\subsubsection{Pattern 11 – Smart Contract Translation}
\paragraph{\textit{Summary}.} Translate smart contract code from one language to another.

\paragraph{\textit{Context}.} Concert application uses a set of smart contracts and associated states which need to be usable on the target blockchain. The smart contract language of the target blockchain is not interoperable. The list of smart contracts and their states to be migrated is given in the snapshot.

\paragraph{\textit{Problem}.} Same as pattern ten.

\paragraph{\textit{Forces}.} Same as pattern ten.

\begin{figure*}[htb!]
	\centering
	\includegraphics[width=\textwidth]{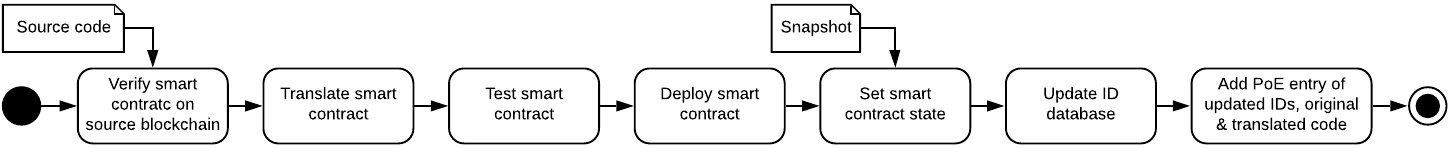}
	\caption{Process of smart contract translation.\label{fig:sc_translation}}
\end{figure*}

\paragraph{\textit{Solution}.} Follow the process outlined in Fig. \ref{fig:sc_translation} to translate and deploy a smart contract on the target blockchain. First, verify that the respective source code produces the exact smart contract deployed on the source blockchain. Second, translate the smart contract to the new language. Then, test the functional correctness and security of the translated contract. Fourth, deploy the new smart contract to the target blockchain using the \textit{state initialization} pattern. Fifth, use the same pattern to set the states of the deployed contract, as per the state recorded on the snapshot. Next, update the mapping between old and new smart contract addresses on the ID database, as smart contract addresses vary across different blockchain instances and could also depend on the address that deployed the smart contract, transaction sequence number, among others. Finally, it is also desirable to include the original and translated code, snapshot file, and mapping between old and new smart contract addresses as a PoE entry, as they are not included in the target blockchain's history.

Smart contracts updated or rewritten for the same blockchain instance are within the purview of this pattern because the new smart contract needs to be redeployed. Further, its states need to be set to the values of the previous contract. When the smart contract language is a compiled one, it is more desirable to compile the source code and then compare it with the binary code on the source blockchain. Compared to syntax-level comparison, this could reveal details such as compiler version and optimizations used during deployment of the original contract, which may need to be taken into consideration during the translation. Depending on the language support and tool availability, translation may happen either at the source or binary code level. In addition to the functional correctness, accuracy of data types, their visibility, and smart contract function-level access control must also be preserved. Also, aspects such as semantic equivalence, event handling, object-oriented features, access control, and optimizations to reduce computation time and execution fees should be taken into account. Ideally, extensive functional and security testing should be performed on the test network provided by the target blockchain. It is not uncommon to use a bug bounty program to test critical smart contracts, e.g., token contracts. Therefore, the migration time frame should allocate sufficient time for such testing. When account owners hold the private keys, users need to create new accounts, deploy new smart contracts, and set their states. Also, if the reference to a smart contract is provided via a smart contract registry \cite{Xu18}, the registry contract needs to be translated and deployed first. Then update the registry with the new smart contract address.

\paragraph{\textit{Consequences}.}
\begin{itemize}
	\item This pattern works for any blockchain platform that has a smart contract language rich enough to recreate the same behavior. However, smart contract translation is not trivial and involves many complications. 
	
	\item It is difficult to guarantee that a translated smart contract exhibits the same behavior. Correctness can be enhanced by verifying the source code before translation, formal verification, and extensive testing.
	
	\item Errors, time, and cost could be reduced via automated translation and testing.		
\end{itemize}

\paragraph{\textit{Related patterns}.} \textit{VM emulation} pattern is preferred when applicable. \textit{State extraction} and \textit{state transformation} patterns can be used to capture the smart contract code, embedded states, aggregate the states, and burn the smart contracts. Use the \textit{state initialization} pattern to deploy and set the state of translated smart contracts. PoE entries of smart contract codes and updates to the contract addresses can be added using the \textit{off-chain data storage} and \textit{encrypting on-chain data} patterns.

\paragraph{\textit{Known use}.} As Serpent was deprecated and had several security weaknesses, Augur \cite{Augur17} smart contract written in Serpent had to be rewritten in Solidity. Kyber \cite{Kyber19} redeployed its contract with fixes, as the contract was about to be unusable due to the revision of Ethereum gas costs. Tronbet \cite{Tronbet19} migrated its ANTE tokens to new WIN tokens managed by a new smart contract with enhanced features. Both Augur and Tronbet used \textit{token burning} and \textit{state initialization} patterns to set the state of the new contract. Counterparty \cite{Counterparty} and Tron \cite{Milano18} allow executing Ethereum smart contracts on Bitcoin and Tron by modifying the Solidity code, respectively. Tools are available to translate among smart contract language, e.g., Ethereum Solidity to Hyperledger Javascript translation \cite{Zafar18}.

\section{Mapping Patterns to Migration Scenarios}\label{mapping}
Given the five migration scenarios, five data fidelity levels, and 13 patterns, the potential solution space is large. Therefore, we present a selected set of use cases from the illustrative example that combines a mix of migration scenarios, data fidelity levels, and blockchain platforms and hosting modes. We then discuss how a migration plan for a given use case could be achieved using a selected set of patterns. Table \ref{table:mapping} shows how the proposed patterns map to the migration use cases.

\begin{table}[!htbp]
	\caption{Pattern to migration scenario mapping.}
	\centering
	\small
	\begin{tabular}{|>{\raggedright}p{33mm}|>{\raggedright}p{38mm}|c|c|c|c|c|}
		\hline
		\multirow{2}{33cm}[-6mm]{\textbf{Pattern Group}} & \multirow{2}{38cm}[-6mm]{\textbf{Pattern}} & \multicolumn{5}{c|}{\textbf{Scenario}} \\ 
		\cline{3-7}
		& & \rotatebox{90}{Relocate} & \rotatebox{90}{Upgrade} & \rotatebox{90}{Consolidate} & \rotatebox{90}{Separation} & \rotatebox{90}{Archive} \\		
		\hline
		State extraction & Snapshotting & \checkmark & \checkmark & \checkmark & \checkmark & \checkmark \\
		\hline		
		\multirow{2}{30mm}{State transformation} & State aggregation & -- & \checkmark & \checkmark & \checkmark & -- \\
		\cline{2-7}
		& Token burning & -- & \checkmark & \checkmark & \checkmark & -- \\
		\hline
		\multirow{5}{30mm}{State and transaction load} & Node sync & ? & -- & -- & ? & ? \\
		\cline{2-7}
		& Establish genesis & ? & ? & -- & ? & ? \\
		\cline{2-7}
		& Hard fork & -- & \checkmark & ? & -- & ? \\
		\cline{2-7}
		& State initialization & \checkmark & \checkmark & \checkmark & \checkmark & \checkmark \\
		\cline{2-7}
		& Exchange transfer & -- & \checkmark & \checkmark & \checkmark & -- \\
		\cline{2-7}
		& Transaction replay & \checkmark & \checkmark & \checkmark & \checkmark & ? \\
		\hline
		\multirow{2}{30mm}{Smart contract} & Virtual machine emulation & -- & ? & ? & -- & -- \\
		\cline{2-7}
		& Smart contract translation & -- & \checkmark & \checkmark & -- & -- \\
		\hline 
		\multirow{3}{30mm}{Non-functional} & Measure migration quality & -- & \checkmark & \checkmark & \checkmark & \checkmark \\ 
		\cline{2-7}
		& Off-chain data storage & -- & \checkmark & \checkmark & \checkmark & \checkmark \\ 
		\cline{2-7}
		& Encrypting on-chain data & -- & \checkmark & \checkmark & \checkmark & \checkmark \\ 
		\hline
		\multicolumn{7}{l}{\footnotesize * Applicable (\checkmark), Maybe applicable (?), Not applicable (--)}			
	\end{tabular}
	\label{table:mapping}
\end{table}

Suppose the nonprofit decided to spin up a private instance of the same blockchain platform and migrate its data. This is an example of the \emph{separation} scenario which can be achieved by spinning up a new blockchain instance and recreating data using most of the migration patterns. \emph{State extraction} and \emph{transformation} patterns, as well as \emph{state initialization} and \emph{establish genesis} patterns can be used to accurately determine, efficiently recreate, and consistently transfer states to the new blockchain instance. \textit{Transaction replay} pattern can be used to establish history. \textit{Hard fork}, \textit{VM emulation}, and \textit{smart contract translation} patterns are not needed, as the target blockchain is new and compatible. Because the blockchain instance is private, the \textit{exchange-transfer} pattern is not desirable due to the cost of using a public exchange. However, the migration tool may act as a decentralized or private exchange that provides a common protocol for state transfer \cite{Sigman, Thomas15}. Similarly, \textit{encrypting on-chain data} pattern is not essential as the target blockchain is private. \textit{Node sync} could also be used where the new nodes can be reconfigured to act as a different blockchain instance after cloning. However, if the private blockchain instance is to be set up using a BaaS, \textit{node sync} and \textit{establish genesis} patterns may not work as the BaaS may not provide node-level access limiting finer control on blockchain software, data, storage, and inter-node communication. Therefore, these patterns are marked as \textit{may be applicable}.

Alternatively, suppose the nonprofit decided to move to a private and incompatible blockchain instance to get better performance and features. In this case, both \textit{separation} and \textit{upgrade} scenarios should be applied together, as only the application-related data are moved to an incompatible blockchain platform. Therefore, the applicable patterns depend on the constraints of each scenario. For example, the \textit{node sync} pattern does not apply as the target blockchain is incompatible, whereas \textit{hard fork} pattern is not needed as the instance is new. \textit{Establish genesis} pattern may not apply when the target instance is hosted as a BaaS. The \textit{VM emulation} pattern is preferred when it is supported by the target blockchain. Else, one can use the \textit{smart contract translation} pattern given that it is possible to translate/rewrite the smart contract to achieve the same functionality.

In another scenario, suppose the nonprofit decided to use an existing but incompatible public blockchain as the target. In this case, both \textit{separation} and \textit{consolidate} scenarios should be applied together. \textit{Node sync}, \textit{establish genesis}, and \textit{hard fork} patterns do not apply, as the target blockchain platform is incompatible, already established, and public. While the \textit{smart contract translation} pattern is applicable, the applicability of \textit{VM emulation} pattern depends on the target blockchain platform. 

Different data fidelity levels could be used under \textit{upgrade}, \textit{consolidate}, and \textit{separation} scenarios. Therefore, the \textit{measure migration quality} pattern should be used in all three scenarios to ensure that the desired subset of data was migrated as per the application and organizational objectives, and respective blockchain constraints are honored. For example, the nonprofit may choose to manage next year's concert budget using data fidelity levels such as \textit{fresh start}, \textit{state only}, or \textit{genesis and transactions}. Moreover, \textit{off-chain data storage} and \textit{encrypting on-chain data} patterns could be used to enhance transparency and privacy, respectively.

If the nonprofit needs to change the blockchain nodes, the \textit{relocate} scenario applies. Relocation could be achieved by cloning a set of new nodes (either on-premise or on the cloud) using the \textit{node sync} pattern. Finally, these nodes can be reconfigured to run as a separate blockchain instance. If the target blockchain is a BaaS instance, \textit{snapshotting} and \textit{state and transaction initialization} patterns could be used to recreate states on new nodes. \textit{Establish genesis} and \textit{hard fork} patterns may not apply when the target instance is hosted as a BaaS. \textit{Smart contract} and \textit{non-functional} patterns are not needed, as the objective is to swap the hardware while enforcing data integrity through blockchain properties. The \textit{archive} scenario applies when the nonprofit needs to keep the full history on a different blockchain instance for data analytics, auditing, or archiving \cite{Paik2019} purposes. As the archive is created on a new blockchain while transferring only a subset of states, transactions, and smart contracts, \textit{hard fork} and \textit{non-functional} patterns also apply based on the chosen data fidelity level, cost, and privacy goals. While these use cases consider migrating only the concert application's data, proposed patterns could also be used to migrate an entire blockchain.

\section{Discussion}\label{discuss}
Next, we briefly discuss practical considerations and challenges in blockchain migration. Migration is relatively straightforward when the target blockchain is a new instance, as it provides greater flexibility in synchronizing or recreating data using \textit{node sync} and \textit{establish genesis} patterns. In contrast, preexisting blockchains, especially the public ones, require extensive efforts to recreate states using relevant patterns. Such efforts involve the use of new transactions to aggregate and initialize states, burn and exchange tokens, disable smart contracts, and redeploy smart contracts, increasing the time and cost of migration. While the \textit{node sync} pattern can be used to establish the history, it can be used only to change the blockchain nodes or when the target blockchain is new and compatible with the source blockchain. It could also fail when the target blockchain is hosted on a BaaS platform. While one could attempt to replay transactions on the target blockchain, it is both costly and time-consuming, unless \textit{state aggregation} style optimizations are used.

Time and cost of migration can be reduced using the \textit{state aggregation} and \textit{transaction replay} patterns depending on the chosen data fidelity level. By lowering the number of states, the risk of failure could also be reduced, which is essential as it is next to impossible to rollback a failed migration on a public blockchain. Therefore, while moving into an existing blockchain, it is desirable to choose a data fidelity level that is sufficient to retain the correct function and availability of the application. The source blockchain can continue to provide access to historical data, if not decommissioned. Otherwise, transactions and block data can be added to the target blockchain as a set of PoE entries using the \textit{off-chain data storage} pattern while substantially reducing the cost and time to migrate. The risk could be further reduced by migrating one state and smart contract at a time, which works with patterns like \textit{state initialization} and \textit{exchange transfer} \cite{Richard18}. Also, by not replaying transactions, we can overcome the issues related to signing them and external accounts. Trial migration on the test network of the target blockchain platform is essential to reduce the risk further \cite{Ruping13}. In conclusion, while the global state can be recreated on any target blockchain with reasonable effort and time, it is desirable not to recreate full history.

Translating smart contracts is not only time and cost-prohibitive but also provides no guarantees on the exact behavior (even with automated translation). While many high-level smart contract languages are still being proposed, they seem to be opting to a few instruction set architectures, e.g., EVM \cite{Hearn16, Buterin17}. Therefore, the reuse of smart contracts at the bytecode level is preferred, as it could better preserve the smart contract behavior. However, code reuse requires both the source and target smart contract execution environment to behave the same way \cite{Amani18}. 

To apply the \textit{token-burning} pattern for smart contracts, it is essential that they implement a self-destruct function before being deployed on the source blockchain. It is recommended to separate the smart contract state from the business logic \cite{Xu18}, as it simplifies smart contract translation and enables efficient migration of smart contract state using patterns such as \textit{snapshotting}, \textit{state aggregation}, \textit{token burning}, \textit{establish genesis}, and \textit{state initialization}. As the addresses could change during the migration, they should not be hardcoded into the smart contract code, e.g., address used in a delegated call. Instead, it is recommended to use a smart contract registry \cite{Xu18} to keep track of smart contract addresses. Similarly, having an ID database at the BAL is essential to handle the change of identifiers during the migration. Else, the migration will require significant changes to the BAL or even the application. Further, the use of states that can be tokenized is useful, as they could be created, aggregated, swapped, and burned using smart contracts. Therefore, proactive system design and application of smart contract best practices are essential to simplify future blockchain migrations.

\section{Summary}\label{summary} While blockchains are designed to be immutable, many technical, business, economic, and regulatory-level changes already making it necessary for an application to migrate from one blockchain instance to another. In this paper, we outlined the need for blockchain migration in DApps, enterprise information systems, and business process management systems. We introduced six new patterns and seven others adapted from the literature to address five migration scenarios and blockchain-specific data management challenges. We further identified five data fidelity levels to balance competing factors such as applications' data requirements, performance, effort, cost, time, security, privacy, and risks of migration failure. While blockchain migration is expected to be difficult and costly, we show that most migration scenarios can be achieved within a reasonable time and cost by choosing a data fidelity level that satisfies the application's minimum requirements and a combination of migration patterns. Some of the challenges requiring further research are: identifying the best data fidelity level for a given application scenario; lack of access to private keys and external accounts; confirming the correctness of translated smart contracts; best practices to simplify future migration; and handling user permissions.
	\bibliography{Ref_new}{}
	\bibliographystyle{ieeetr}
	
\end{document}